\documentclass[letter,scriptaddress,twocolumn, prl,showkeys,showpacs,superscriptaddress,longbibliography]{revtex4-1}

	\usepackage{amsmath}
	\usepackage{makeidx}
	\usepackage{algorithm,algcompatible}
\algnewcommand\INPUT{\item[\textbf{Input:}]}%
\algnewcommand\OUTPUT{\item[\textbf{Output:}]}%
	\usepackage{amsfonts}
	\usepackage{graphicx, xcolor}  
	\usepackage{dcolumn}   
	\usepackage{bm}        
	\usepackage{amssymb}   
	\usepackage[ansinew]{inputenc}
	\usepackage[usenames,dvipsnames]{pstricks}
	\usepackage{multirow}
	\usepackage{caption}
	\usepackage{epstopdf}
	\usepackage{pst-grad} 
	\usepackage{pst-plot} 
	\usepackage{bbm} 
	
\newcommand{\be}{\begin{equation}}
\newcommand{\ee}{\end{equation}}
\newcommand{\ba}{\begin{aligned}}
\newcommand{\ea}{\end{aligned}}
\newcommand{\lp}{\left(}
\newcommand{\rp}{\right)}
\newcommand{\rd}{{\rm d}}

\newcommand{\tr}{\mathop{\mathrm{tr}}}

\newcommand{\ms}{\mathcal{S}}

\newcommand{\mm}{\mathcal{M}}

\begin{document}

\title{Principles for  optimal cooperativity in allosteric materials}
\author{L. Yan}
\thanks{These two authors contributed equally to this work}
\affiliation{Kavli Institute for Theoretical Physics, University of California, Santa Barbara, CA 93106, USA}
\author{R. Ravasio}
\thanks{These two authors contributed equally to this work}
\affiliation{Institute of Physics, \'Ecole Polytechnique F\'ed\'erale de Lausanne, CH-1015 Lausanne, Switzerland}
\author{C. Brito} 
\affiliation{Instituto de F\'isica, Universidade Federal do Rio Grande do Sul, CP 15051, 91501-970 Porto Alegre RS, Brazil}
\author{M. Wyart}
\email{matthieu.wyart@epfl.ch}
\affiliation{Institute of Physics, \'Ecole Polytechnique F\'ed\'erale de Lausanne, CH-1015 Lausanne, Switzerland}

\begin{abstract}
Allosteric proteins transmit a mechanical signal induced by binding a ligand. However, understanding  the nature of the information transmitted and the architectures optimizing such transmission remains a challenge. Here we show using an {\it in-silico} evolution scheme and theoretical arguments that  architectures optimized to be cooperative, which propagate efficiently  energy, {qualitatively} differ from previously investigated materials optimized to propagate strain. Although we observe a large diversity of functioning cooperative architectures (including shear, hinge and twist designs), they all obey the same principle {of displaying a {\it mechanism}, i.e. an  extended {soft} mode}. We show that  its optimal frequency decreases with the spatial extension $L$ of the system as $L^{-d/2}$, where  $d$ is the spatial dimension. For these optimal designs, cooperativity  decays logarithmically with  $L$ for $d=2$ and does not decay for $d=3$. Overall our approach leads to a natural explanation for several observations in allosteric proteins, and { indicates an experimental  path to test if allosteric proteins lie close to optimality}.
\end{abstract}

\maketitle

\section*{Introduction}

Many proteins are {\it allosteric}: binding a ligand at an allosteric site can affect the properties of a distant active site, sometimes located on the other side of the protein \cite{Monod65,Changeux05}. Predicting the existence of such allosteric pathways from  protein structure alone would be of great interest \cite{Amor16,Halabi09}, since they can be used as targets for drug design \cite{Nussinov13}.  Solving this challenge  requires to make progress on   both physical and biological questions. First, how can such {disordered materials \cite{Liang01}} be designed to carry mechanical information specifically over long distances? Are there fundamental limits to what can be achieved? Second, what are allosteric pathways really optimized for? What kind of elastic information do they carry?  A physical theory of allostery should address these points. It should also explain  the following empirical facts:  (i) Some allosteric proteins \cite{Gerstein94}, including hemoglobin \cite{Perutz70,Xu03}, essentially function as  hinges, while others display a ``shear''  design where  two  rigid parts are connected by a weak plane \cite{Mitchell16}.  This classification is however not exhaustive, as in various cases the response  to binding  a ligand cannot be described in term of a simple shear or hinge motion  \cite{Goodey08,Gandhi08,mclaughlin12}. (ii) The response to binding often  corresponds mostly to motion along few soft normal modes of the protein \cite{Atilgan01,Rios05}. These modes tend to be conserved during evolution \cite{Zheng06}. (iii) In some cases the allosteric functional effect at the active site is significant while the physical mean displacement  induced by binding the ligand is small. It has been proposed  that for these proteins binding can affect how particles near the active site fluctuate around their mean position, while changing little the latter {\cite{Popovych06,Cooper84,Tsai08,McLeish13}}.  

Recently, allostery  was investigated  using {\it in silico} evolution schemes where a system evolves to perform a given function \cite{Hemery15,Tlusty16,Yan17,Rocks17,Flechsig17}. Most relevant here are schemes developed to solve  inverse elastic problems \cite{Yan17,Rocks17,Flechsig17},  {in a spirit similar to topology optimization used in engineering to design  functional tools from compliant materials \cite{Sigmund13,Sigmund97,Nishiwaki98}. The  task studied in \cite{Yan17,Rocks17,Flechsig17} was to }design a material whose response to a specific local strain applied on one of its sides (the allosteric site) leads to a displacement whose {{\it geometry}} is prescribed on the opposite side (the active site). Under broad conditions these algorithms find solutions that achieve such ``geometric'' tasks essentially perfectly. The corresponding architectures turn out to have {surprising properties: their} response almost vanishes in the bulk of the material and reappears near the active site \cite{Yan17}. This amplification of the elastic signal is caused by the emergence of a powerful lever, made of a soft elastic region surrounding the active site, where the system is just constrained enough to act as a solid \cite{Yan17a,Yan17}.  Although {there is great interest in finding whether such} architectures exist in nature, an intriguing aspect of this approach is that it does not generate the well-known  { allosteric architectures} such as {the simple shear and hinge designs, in which the response  remains of similar magnitude between the allosteric and the active sites}. 

Here we show that a simple modification of the task, where materials are optimized such that the binding at the allosteric site lowers {\it the binding energy} of another ligand at the active site, leads to  different design principles. {
In the context of proteins, this task corresponds to maximizing the cooperativity of binding two ligands, a central feature of various allosteric proteins \cite{Monod65}.
We find that} there is a zoology of architectures achieving such cooperativity, but they always display a stiff structure (embedded in a softer elastic matrix) with a single very soft extended elastic mode or ``mechanism". We lay out the  principles behind such designs, and show theoretically that the soft mode frequency should be neither too large nor too small to optimize function: {its optimal value decreases with the material size, and scales as $L^{-d/2}$ in spatial dimension $d$}. We prove that cooperativity then decays as $\ln^{-1}(L/c)$ for $d=2$ and is even independent of $L$ in larger spatial dimensions $d\geq 3$, where $L$ is the linear extension of the system and $c$ the length scale on which binding takes place. This result is very different from a normal continuous elastic medium where cooperativity rapidly decays with distance as $L^{-d}$. 
Overall the classification we provide  leads to  a natural explanation for the key aspects of allostery described in (i,ii,iii). It also shows that a path of large strain values connecting the allosteric and active site induced by binding is not necessary for cooperativity to occur, and it makes further testable predictions, including the locations where a shear or hinge design would be mostly affected by a mutation and conserved during evolution.

\begin{figure}[htbp!]
\includegraphics[width=1.\columnwidth]{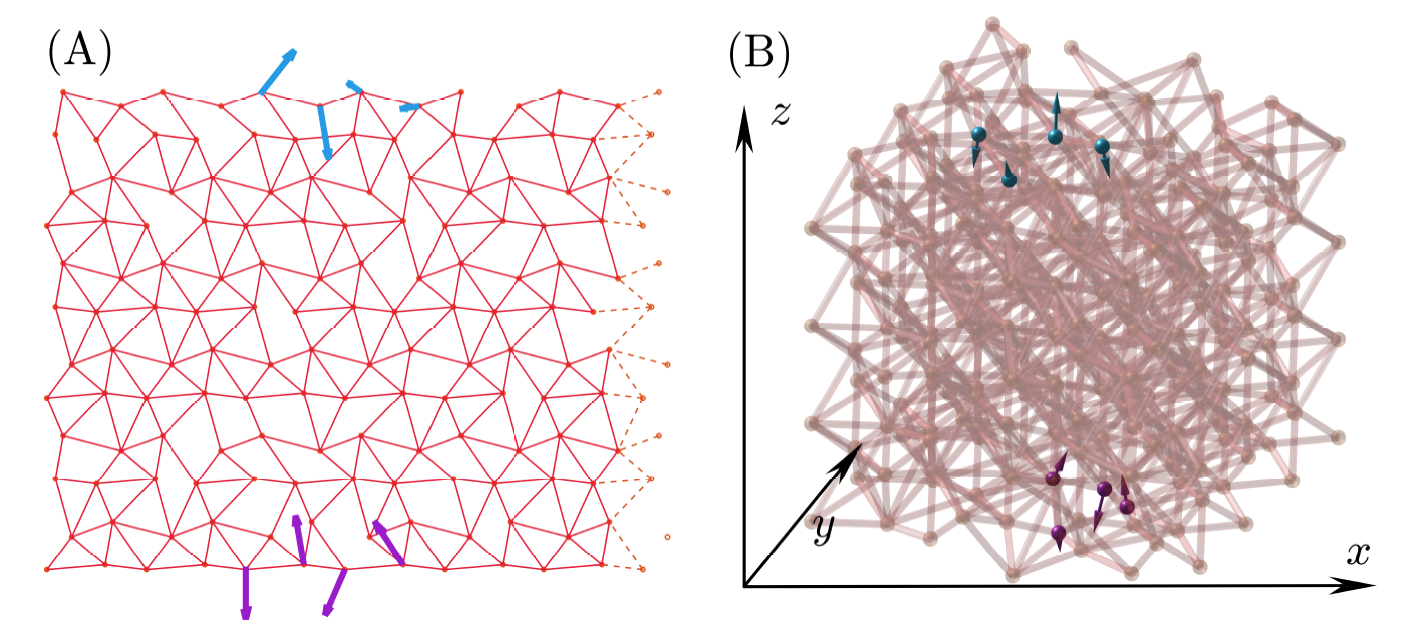}
\caption{\small{Examples of on-lattice elastic networks. (A) shows a hexagonal lattice ($d=2$) with 
periodic boundary conditions along the horizontal axis (springs crossing the periodic boundary are shown in dashed lines, and are not present when open boundary conditions are used), mimicking a cylindrical geometry. (B) For $d=3$, we use a face centered cubic lattice 
with open boundaries. In all cases, occupied links displaying a spring of stiffness unity are indicated by lines. The stimulus displacement is shown in purple arrows and the target displacement is shown in blue arrows, each are applied on four nodes. All data are presented for $L=20$ and $z=5.0$ in $d=2$ and $L=12$ and  $z=8.4$ in $d=3$.
}}\label{fig_model}
\end{figure}

\section*{Methods}

\subsection{ In-silico Evolution Scheme}

{\bf Elastic networks:}
To model allosteric materials we consider elastic networks, often used to describe proteins~\cite{Atilgan01,Rios05,Zheng06}.  
Specifically, $N=L^d$ nodes are located on a lattice (slightly distorted periodically to avoid straight lines as discussed in Supplemental Material Section A and \cite{Yan14,Yan15a}), and  among all $N_c$ links of nearest nodes, a subset of $N_{\rm s}$ pairs are connected 
  by harmonic springs of stiffness  $k=1$ , as indicated by lines in Fig.~\ref{fig_model}. We declare that  $\sigma_{\alpha}=1$ if a spring is present in the link $\alpha$ and $\sigma_{\alpha}=0$ otherwise. Thus the network is entirely described by a connection vector $|\sigma\rangle$ made of zeros and ones, whose dimension is the number of links {$N_c$}. We define the average coordination number $z\equiv 2 N_{\rm s}/N$ { and average connection $\bar{\sigma}=N_{\rm s}/N_c$} and keep {them} fixed during evolution. We find that our results do not depend qualitatively on $z$ as long as $z> z_c=2d$, the rigidity limit derived by Maxwell ~\cite{Maxwell64}.

{\bf Binding:} Binding a ligand exerts forces locally that leads to an imposed local strain. To model this effect at the allosteric site, we choose four adjacent nodes on one side of the system (shown in purple in Fig.~\ref{fig_model}), and consider that binding at that site imposes a displacement $|\delta{\bf R}^{{\cal A}l}\rangle$ on these nodes, as indicated by purple arrows. {(Strictly speaking, this description of binding assumes that the ligands  are rigid. However we expect our results to hold true qualitatively as long as the ligands are not significantly softer than the protein itself).} Minimizing the elastic energy in the entire system with these constraints then leads to a response  $|\delta{\bf R(\sigma)}_r^{{\cal A}l}\rangle$ that can be extended  (see a formal expression for this response in {Supplemental Material Section B} and~\cite{Yan17}). 
The corresponding energy cost associated with binding writes:
\be
E^{{\cal A}l}(\sigma) = \frac{1}{2}\langle \delta{\bf R}_r^{{\cal A}l}|{\cal M}|\delta{\bf R}_r^{{\cal A}l}\rangle,
\label{eq_energy}
\ee
where ${\cal M}$ is the stiffness matrix of the network (whose definition is recalled in {Supplemental Material Section B}) of dimension $Nd\times Nd$, which depends on the network considered.  The same procedure is used to model the binding of another ligand at the active site (indicated in blue in Fig.~\ref{fig_model}), allowing us to define a binding energy $E^{{\cal A}c}(\sigma)$. If the two binding events take place simultaneously, the same procedure leads to the derivation of a joint binding energy $E^{{\cal A}c, {\cal A}l}(\sigma)$. 

\begin{figure}[htbp!]
\centering
\includegraphics[width=0.8\columnwidth]{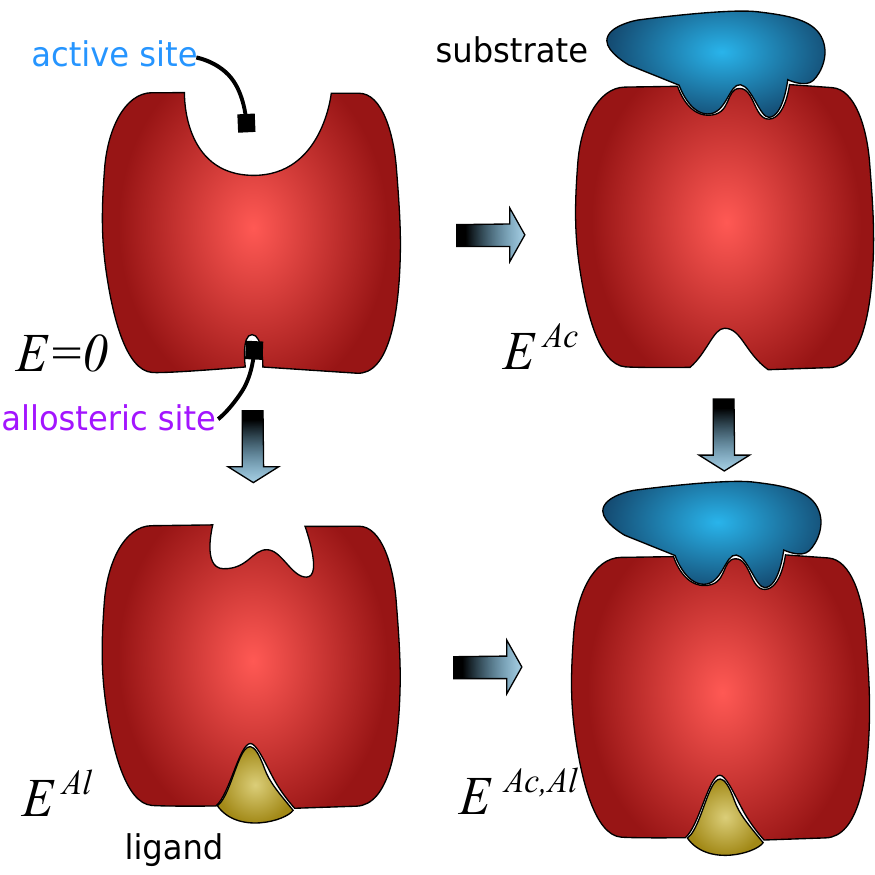}
\caption{\small{Illustration of cooperativity. With two binding sites, a protein displays four states. Cooperativity is high if binding  a substrate molecule at its active site  is difficult when the allosteric site is empty (i.e. $E^{{\cal A}c}$ is large) whereas it is much simpler when the allosteric site is occupied (i.e. $E^{{\cal A}c,{\cal A}l}-E^{{\cal A}l}$ is small).}
}\label{fig_cooperative}
\end{figure}

{\bf Cooperativity:} {W}e seek to engineer materials in which binding at the allosteric site lowers  the  binding energy at the active site as much as possible, as illustrated in Fig.~\ref{fig_cooperative}. In the absence of the ligand at the allosteric site, the binding energy at the active site is simply 
$E^{{\cal A}c}(\sigma)$, whereas if present  it is $E^{{\cal A}c, {\cal A}l}(\sigma)- E^{{\cal A}l}(\sigma)$. We seek to maximize the cooperative energy, simply defined as the difference between these terms: 
{ 
\be
E_{\rm coop}=E^{{\cal A}c}(\sigma)+ E^{{\cal A}l}(\sigma)-E^{{\cal A}c, {\cal A}l}(\sigma)\equiv{\cal F},
\ee
which also defines our fitness function.
}

Cooperativity  turns out to differ greatly from the geometric task in which a displacement imposed at one end of the material must elicit a given displacement at the other end \cite{Yan17,Rocks17,Flechsig17} {(see below and Supplemental Material for a detailed comparison)}. The architectures  associated with the latter task are very asymmetric, in particular they are much softer near the active site than near the allosteric site \cite{Yan17}. By contrast, it is clear from our definition of cooperativity that both active and allosteric sites play a symmetric role. 
At an intuitive level, the difference  can be understood by considering the limit of weak elastic coupling between allosteric and active sites for which one finds ${ E_{\rm coop}}\approx \langle F^{{\cal A}c}|\delta R^{{\cal A}l\rightarrow {\cal A}c}\rangle$ where $|F^{{\cal A}c}\rangle$ is the external force field generated by  the substrate when it binds to the active site, and $|\delta R^{{\cal A}l\rightarrow {\cal A}c}\rangle$ is the displacement field induced at the active site by binding a ligand at the  allosteric site. 
Maximizing cooperativity thus requires to have a large  and specific response  $|\delta R^{{\cal A}l\rightarrow {\cal A}c}\rangle$ 
(which is essentially what the geometric task accomplishes) {\it and} to have a large force scale $|F^{{\cal A}c}\rangle$, which requires the material to be stiff near the active site. This additional constraint makes the cooperative task  harder than the geometric one.

\begin{figure}[htbp!]
\includegraphics[width=0.9\columnwidth]{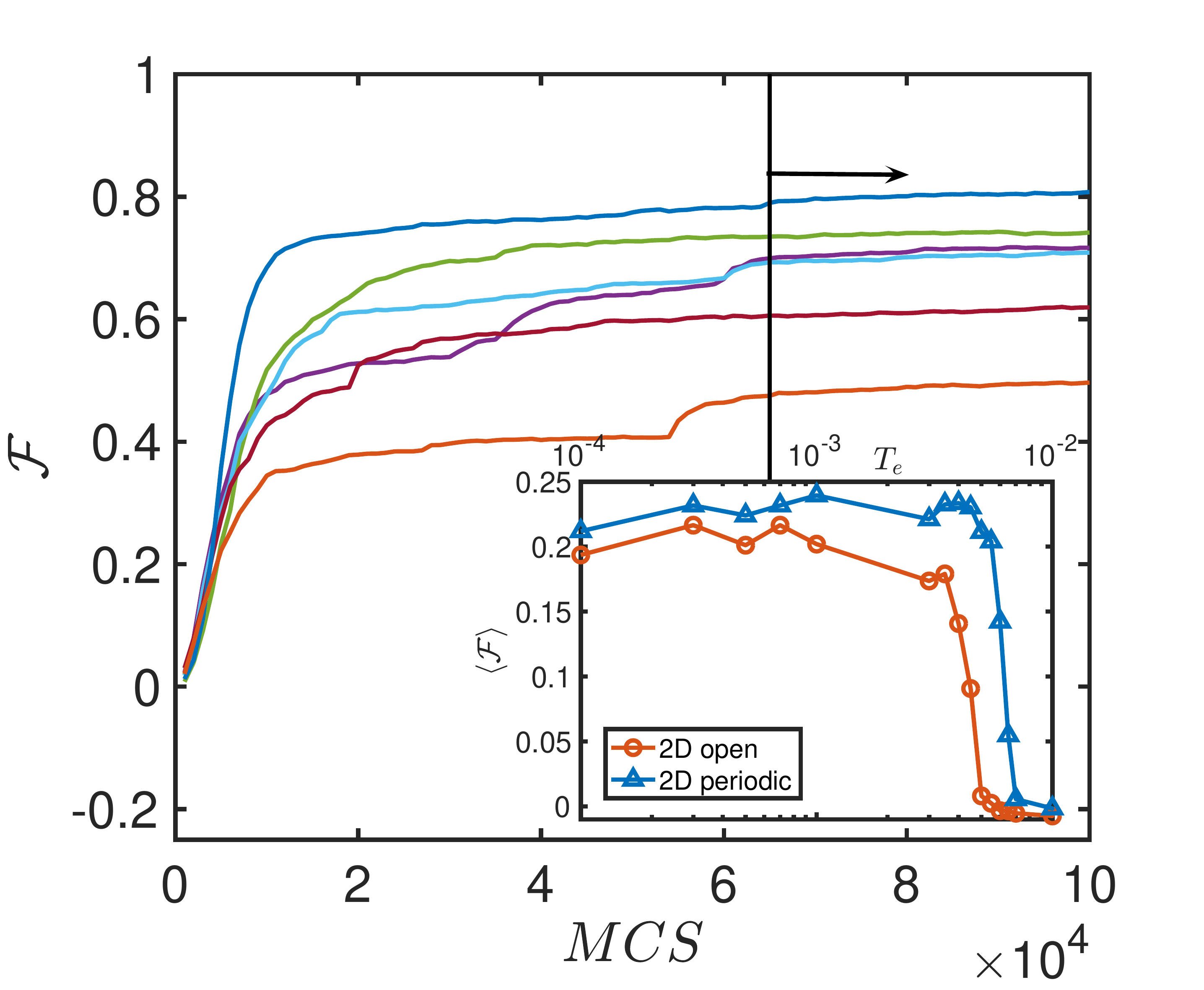}
\caption{\small{Evolution of the {fitness} ${\cal F}$ {\it vs} the number of Monte Carlo steps $MCS$. Different initial conditions resulted in different architectures, which are analyzed at sufficiently long time to avoid significant transient effects (keeping only the { data from the last $3.5\times10^4$ steps out of the $10^5$ $MCS$ in each run}, as delimited by the black vertical line in the plot). {The inset shows the fitness ${\cal F}$ averaged over 25 initial conditions as a function of the evolution temperature $T_e$ for the two dimensional network with both open and periodic boundaries}. }
}\label{fig_lineage}
\end{figure}

{\bf Evolutionary Dynamics:}
To generate cooperative architectures, we implement an  evolution scheme which selects preferably networks with high fitness. Specifically, we use a Monte-Carlo algorithm where the relocation of individual springs is considered, i.e. $|\sigma\rangle\to|\sigma'\rangle$ where a randomly chosen vacant link $\gamma$  becomes occupied $\sigma_\gamma=0\to\sigma'_\gamma=1$ and a randomly occupied link $\alpha$ becomes empty, $\sigma_\alpha=1\to\sigma'_\alpha=0$. The new structure is selected with the probability $p=\min[1,\exp(\frac{{\cal F}(\sigma')-{\cal F}(\sigma)}{T_e})]$, where $1/T_e$ is inverse evolutionary ``temperature'' characterizing the selection pressure. 

We find that as the selection pressures increases and $T_e$ decreases,  there is a rather sudden  transition from non-working networks with zero fitness to cooperative ones, as illustrated in inset of Fig.\ref{fig_lineage}. The fitness then appears to plateau,
and in what follows we choose $T_e=10^{-4}$ where this plateau is reached.  Interestingly, in this plateau region we find that the fitness landscape is glassy: 
there are many {\textit{families}} of solutions that are not dynamically connected on the time scale of our runs, implying the presence of large fitness barriers. The families obtained in a given run {are defined by the respective} initial conditions, and do not display exactly the same fitness as shown in  Fig.\ref{fig_lineage}. We checked that sequences are much more similar within a family than between different families. Indeed, in a single family the mean overlap between  distinct configurations $i$ and $j$, $q\equiv \langle \overline{\sigma^i_\alpha \sigma^j_\alpha}^{\alpha}\rangle- \overline\sigma^2$ is high with $q\approx0.36$, while it is small $q\approx 0.{0}3$ for different  families ($\overline{\bullet}^\alpha$ averages over links, and $\langle\bullet\rangle$ averages over configurations). Glassiness also implies that the architectures slowly evolve in time, but less and less so as time goes on. In what follows, we study architectures only in the last third of the run,   when transient effects are weaker and fitness is nearly stationary.
{In total, we generated 25 families in $d=2$ and 10 families in $d=3$.}

\subsection{Analysis Toolbox}
In this section we review useful observables characterizing allosteric architectures. Most of them are known in the protein literature, others are novel to the best of our knowledge. 

{\bf Geometry of  allosteric response:} 
By computing the structure of proteins crystallized with and without the ligand bound on their allosteric site, 
one gets access to the internal response of the protein induced by binding, $|\delta{\bf R}_r^{{\cal A}l}\rangle$ in our notations. 
As recently emphasized in this context \cite{Mitchell16}, a key aspect of this response is its strain, which must be zero in parts of the proteins moving as rigid blocks. The strain thus captures where deformation is actually taking place. 
The strain tensor $\overset\leftrightarrow{\epsilon}(i)$ can be directly computed from any displacement $|\delta{\bf R}\rangle=\{\delta R_i\}$ where $i$ labels particles or nodes, as shown  in {Supplemental Material Section C} or Ref.~\cite{Gullett07}. Removing the trace leads to a local shear tensor $\overset\leftrightarrow{\gamma}(i) = \overset\leftrightarrow{\epsilon}(i)-\frac{1}{d}\tr[\overset\leftrightarrow{\epsilon}(i)]\mathbbm{1}$, where $\mathbbm{1}$ is a $d\times d$ identity matrix.  It's useful to define scalar observables to visualize the strain, in particular the shear intensity $E_{\rm shear}(i)$ (not sensitive to compression or dilation) and the bulk intensity $E_{\rm bulk}(i)$ (sensitive to it) as ~\cite{Mitchell16}:
\be\ba
E_{\rm shear}(i) &= \frac{1}{2}\sum_{l,m=1}^{d}[\gamma_{lm}(i)]^2;\\
E_{\rm bulk}(i) &= \frac{1}{2}\sum_{l=1}^{d}[\epsilon_{ll}(i)]^2.
\label{eq_shbk}
\ea\ee

{\bf Rigidity of the structure:} For elastic networks, as understood by Maxwell an important aspect of  rigidity is the coordination number $z(i)$, counting the local connectivity (number of springs) attached to a node $i$. This notion, sufficient in our model,  can be extended to interactions relevant in proteins as discussed in \cite{Jacobs01}.

Another commonly used observable  is the B-factor or Debye-Waller factor~\cite{Yang09}. It characterizes the mean square thermal fluctuations of the particle positions. In a harmonic approximation it can be expressed in terms of the vibrational modes (neglecting a temperature-dependent pre-factor):
\be
B(i) = \sum_{\omega>0}\frac{1}{\omega^2}\delta{\bf R}_\omega(i)\cdot\delta{\bf R}_{\omega}(i),
\label{eq_bfac}
\ee
where the $\omega$s and $\delta{\bf R}_\omega$ are  frequencies and the corresponding vibrational modes, that are obtained from the diagonalisation of the stiffness matrix.

\begin{figure*}[htbp]
\includegraphics[width=\linewidth]{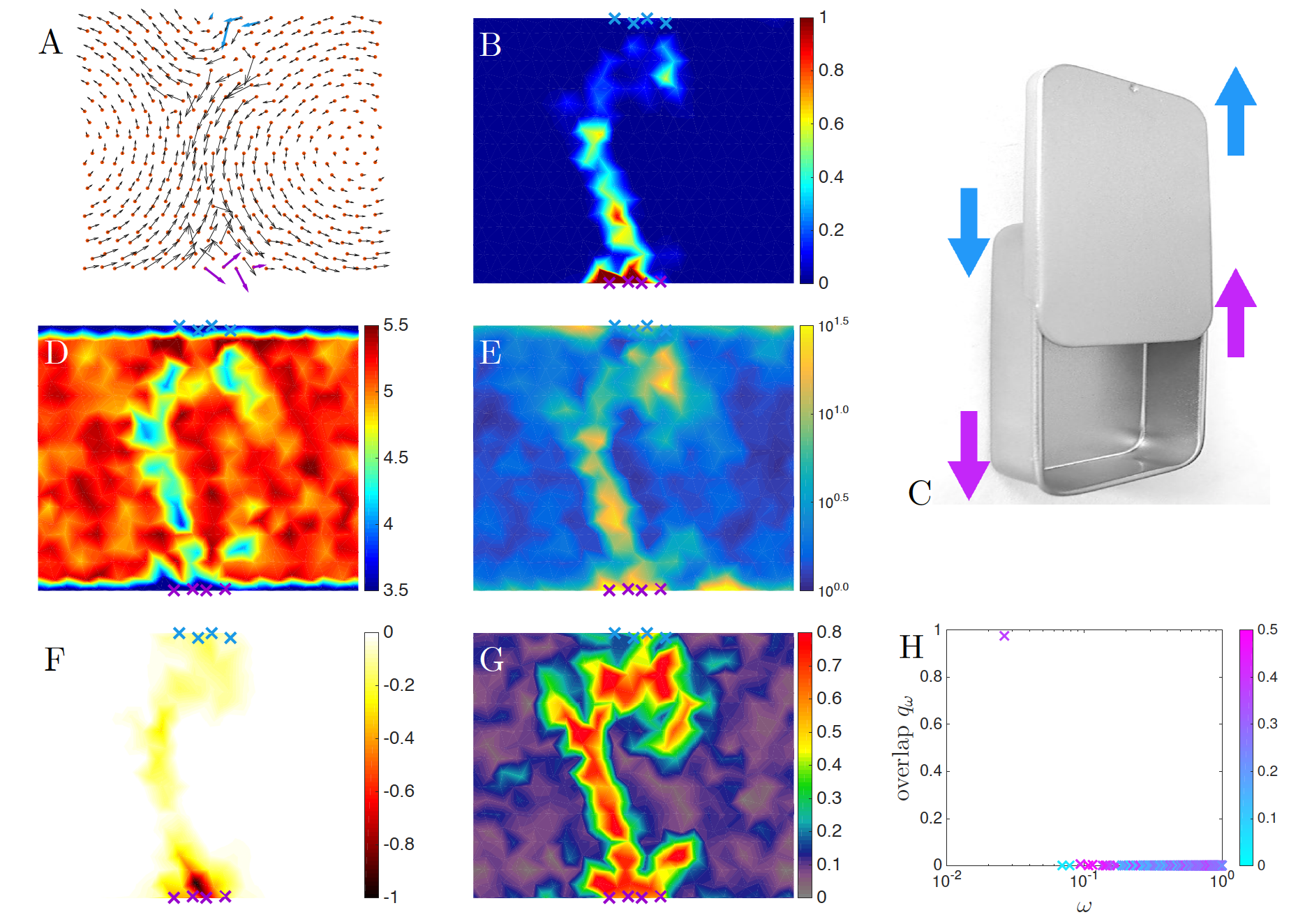}
\caption{\small{{ Shear design:} (A) The average cooperative response $\delta{\bf R}_r^{{\cal A}l}$ induced by binding at the allosteric site  is shown in black arrows. (B) The average shear  intensity  map $E_{\rm shear}$ reveals strain localization along a path. (C) A mint box that opens by sliding  illustrates the shear mechanism.  (D) Map of the average coordination number $z$. (E) Map of the average strain B-factor $SB$. (F) Map of the { fitness cost of single site mutation normalized by its absolute value} $\Delta{\cal F}/{\cal F}$. (G) Map of the conservation $\Sigma$ in the evolution simulation. (H) Overlap $q_\omega$ between the response and the vibrational modes, colored as a function of the participation ratio $P_\omega$, showing that a single extended mode dominates the response to binding. }
}\label{cooppb}
\end{figure*}

B-factors however may not pick up the interesting flexibility of the structure. For example if a hinge connects two rigid parts,  B-factors may be large in the rigid parts too as it is sensitive to rigid motions as well. Here we introduce an observable that would reveal the presence of a hinge, as it characterizes the   thermal fluctuations of the strain (which is therefore zero by construction for rigid body).
We call it the strain B-factor, which for harmonic dynamics follows:
\be
SB(i) = \sum_{\omega>0}\frac{2}{\omega^2}[E_{\rm shear,\omega}(i)+E_{\rm bulk,\omega}(i)],
\label{eq_sbfac}
\ee
where $E_{\rm shear,\omega}$ and $E_{\rm bulk,\omega}$ are the shear and bulk intensities for a given mode $\delta{\bf R}_\omega(i)$, as defined from Eqs.(\ref{eq_shbk}). 

{\bf Spectral analysis:} The response to binding can be decomposed into the vibrational modes ~\cite{Zheng06}, which form a complete orthogonal basis. We define the overlap:
\be
q_\omega = ||{\langle\delta{\bf R}_r^{{\cal A}l}}|\delta {\bf R}_\omega\rangle||^2/||{\delta{\bf R}_r^{{\cal A}l}}||^2,
\label{eq_overlapq}
\ee
that satisfies $\sum_\omega  q_\omega =1$.

The extendedness of the vibrational modes is characterized by  the participation ratio, defined as:
\be
P_\omega=\lp N\sum_i(\delta{\bf R}_\omega(i)\cdot\delta{\bf R}_\omega(i))^2\rp^{-1}
\label{eq_partratio}
\ee
for normalized modes $\sum_i\delta{\bf R}_\omega(i)^2=1$. Translations have a unity participation ratio. By contrast, if a mode only involves the motion of $\sim N_0$ particles, then $P_\omega\sim N_0/N$.


{\bf Conservation:} We quantify the local conservation of the structure by considering the mean occupancy, defined over a period of observation $\tau$:
\be
{\langle\sigma_\alpha\rangle}\equiv\frac{1}{\tau}\sum_{t=1}^{\tau}\sigma_\alpha(t).
\label{eq_sigma}
\ee
If there is no selection pressure on that link, we expect {$\bar\sigma$}. We thus define the conservation $\Sigma$ to quantify the deviation from this average~\cite{Yan17}:
\be
\Sigma_\alpha=\langle\sigma_\alpha\rangle\ln\frac{\langle\sigma_\alpha\rangle}{\bar\sigma}+(1-\langle\sigma_\alpha\rangle)\ln\frac{1-\langle\sigma_\alpha\rangle}{1-\bar\sigma}.
\label{eq_conserva}
\ee

\section*{Results}

We now document examples of architectures generated by our scheme, focusing on shear, hinge and twist designs.
We consider individual families: when average quantities are presented, they always correspond to a time average over the last third of our Monte-Carlo algorithm,  as previously described. We then emphasize the features common to all these designs, to be explained in the next section.

\begin{figure*}[htbp]
\includegraphics[width=1\linewidth]{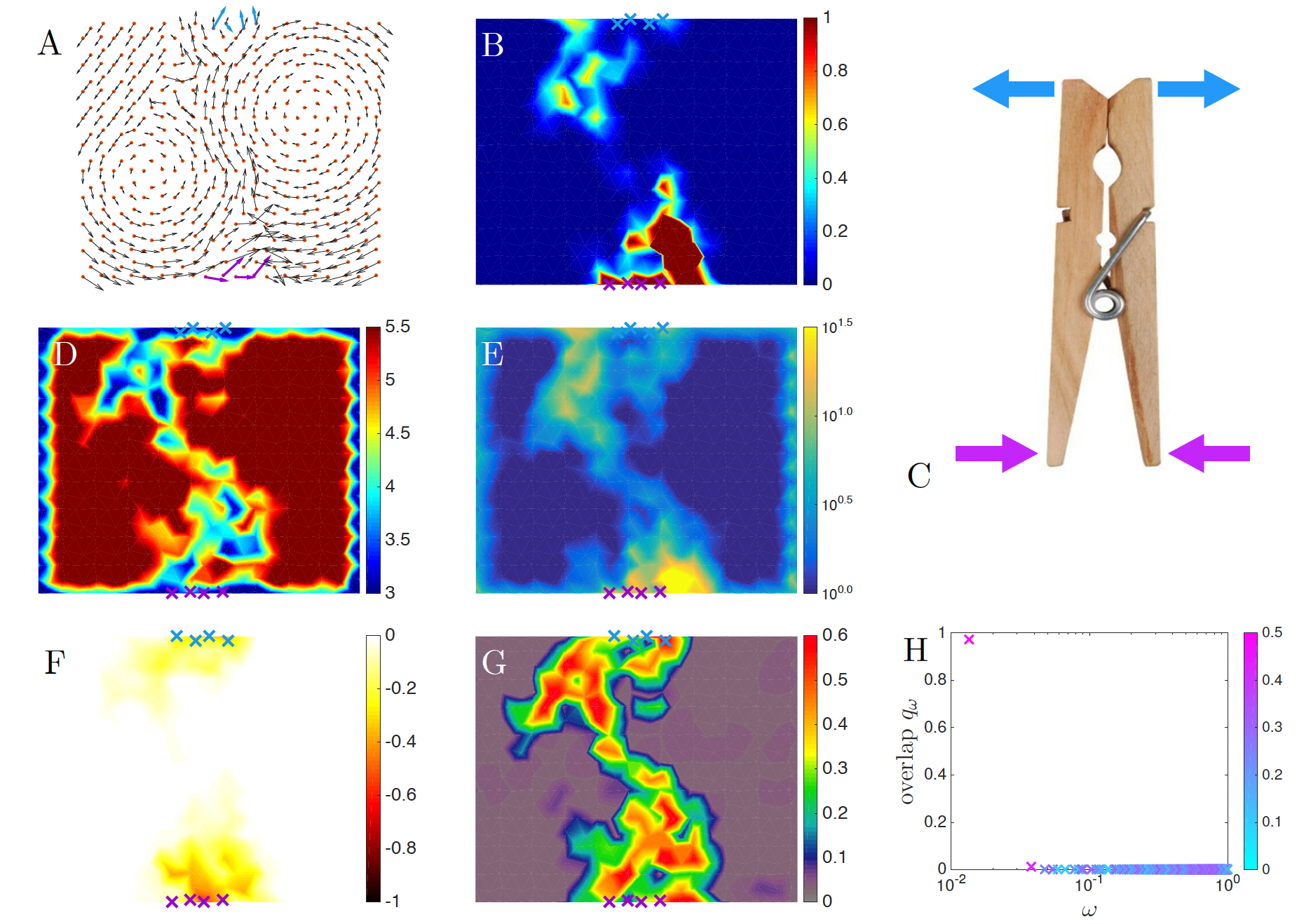}
\caption{\small{{ Hinge design:} (A) The averaged cooperative response $\delta{\bf R}_r^{{\cal A}l}$ induced by binding at the allosteric site  is shown in black arrows. (B) Shear  intensity  $E_{\rm shear}$ of the response. (C) A clothespin  illustrates the hinge mechanism.  (D) Map of the average coordination number $z$. (E) Map of the average strain B-factor $SB$. (F) Map of the { fitness cost of single site mutation normalized by its absolute value} $\Delta {\cal F}/{\cal F}$. (G) Conservation $\Sigma$. (H) Decomposition  $q_\omega$ of the response on the vibrational modes $\omega$, colored as a function of the participation ratio $P_\omega$.
}
}\label{coopob}
\end{figure*}

{\bf Shear design:} We start by the two-dimensional case where visualization is  easier. If periodic boundary conditions are considered on the horizontal axis (cylindrical geometry), we find that {all 25} architectures correspond to a shear design. 
This is illustrated in Fig.~\ref{cooppb}A showing the response to binding: except for a linear path connecting the allosteric and active sites, the motion is essentially that of a rigid body (pure rotations and translations). This is most obvious when plotting the map of the shear intensity $E_{shear}$  in Fig.~\ref{cooppb}B, which is essentially zero excepted along that path. Overall, the design is similar to that of the mint box illustrated in Fig.~\ref{cooppb}C, where strain also localizes on a hyperplane  (a line for $d=2$ and a plane fr $d=3$). { At the}  structural level, we find that the strain path corresponds to a softer region with lower coordination as shown in Fig.~\ref{cooppb}D  and a larger  strain B-factor, as illustrated in Fig.~\ref{cooppb}E.

{\bf Hinge design.} When open boundaries (instead of periodic ones) are used, we find that {about 40 to 50 percent} of the families lead to hinge architectures, {and the rest display a shear design}. In the {former} case, the response exemplified in Fig.~\ref{coopob}A can be decomposed into the motion of two rigid bodies connected by a hinge.  Again this is most apparent in the map of the shear intensity  in Fig.~\ref{coopob}B, showing that there is little strain excepted for two disconnected regions near the allosteric and active sites. There is thus no connecting path of high strain between these sites.  This design is common in our daily life, as illustrated by the clothespin  in Fig.~\ref{coopob}C. At the structural level, the map of coordination shown in Fig.~\ref{coopob}D and that of strain B factor shown in Fig.~\ref{coopob}E display a ``H'' shape with two rather disconnected region being weakly coordinated with a high strain B-factor.

\begin{figure*}[htbp]
\includegraphics[width=\linewidth]{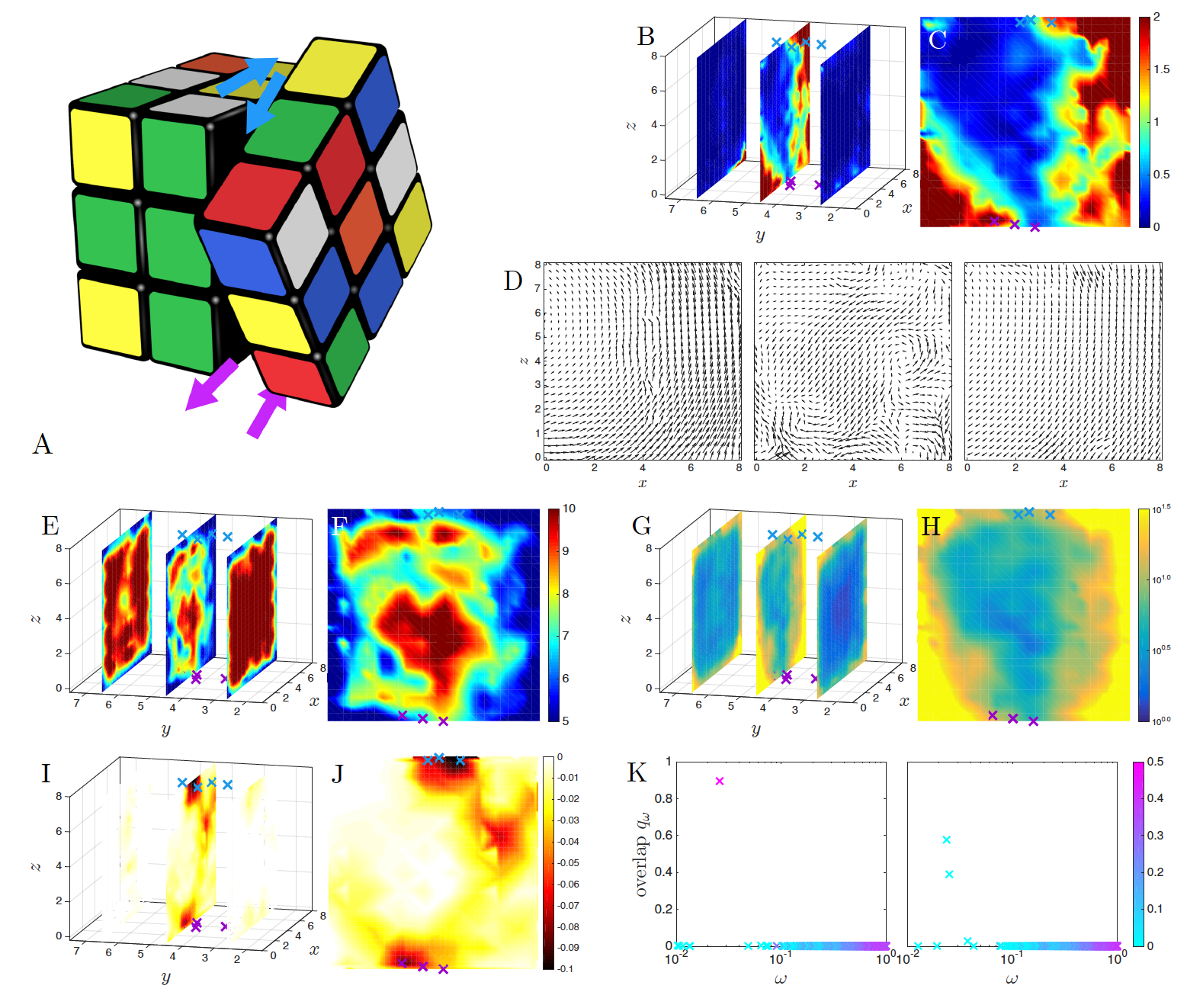}
\caption{\small{{ Twist design:} (A) Illustration of a Rubik Cube and its twist mechanism. (B) {Two-dimensional sections} of  the shear strain intensity.  The allosteric and active sites are shown in purple and blue respectively. (C) Shear intensity  $E_{\rm shear}$  in the central section. (D) Response $\delta{\bf R}_r^{{\cal A}l}$ to binding in the same three distinct sections, organized from left to right as in (B).  Maps of the average coordination number $z$ on (E) the three sections and (F) the central section. The strain B-factor $SB$ is shown on  the three sections (G) and  the central one (H).  { Fitness cost of single site mutation normalized by its absolute value} $\Delta {\cal F}/{\cal F}$ on (I) the three sections and (J) the central one. (K) Decomposition  $q_\omega$ of the response on the vibrational modes  {\it vs}  the mode frequency ${\omega}$, colored  as a function of their participation ratios $P_\omega$, at two different time points during the run. In three dimensions, most of the spectral decomposition resembles the right panel where several vibrational modes project on the response, although we can always identify time points where a single mode contributes as shown on the left panel.}
}\label{coop3D}
\end{figure*}

{\bf Twist design.} In three dimensions  we find a rich variety of architectures,  whose structure and response are sometimes hard to describe. Here we present the simple case of a twist architecture, as illustrated in Fig.~\ref{coop3D}A with the Rubik Cube. 
To visualize this design, we consider the shear intensity in three  sections  parallel to the $x$-$z$ plane as illustrated in  Fig.~\ref{coop3D}B. We find that there is little strain  except on the central plane connecting the allosteric (purple) and active (blue) sites shown in  Fig.~\ref{coop3D}C. There is not however a homogeneous shear on that plane: instead,  the strain is low at its center and larger near the boundaries. Further evidence for the twist design appears in the allosteric response itself shown in Fig.~\ref{coop3D}D with the same slicing geometry: the two side planes show  reverse rotating motions, whereas the middle plane shows a more complex displacement pattern. Once again, the structural analysis confirms this view: we find that the coordination is large and the strain B-factor is small except near  the boundaries of the central plane, as shown in Figs.~\ref{coop3D}(E-H). The middle of the central plane thus acts as a well-connected joint around which two quite rigid bodies can rotate.

\begin{figure}[htbp]
\includegraphics[width=\columnwidth]{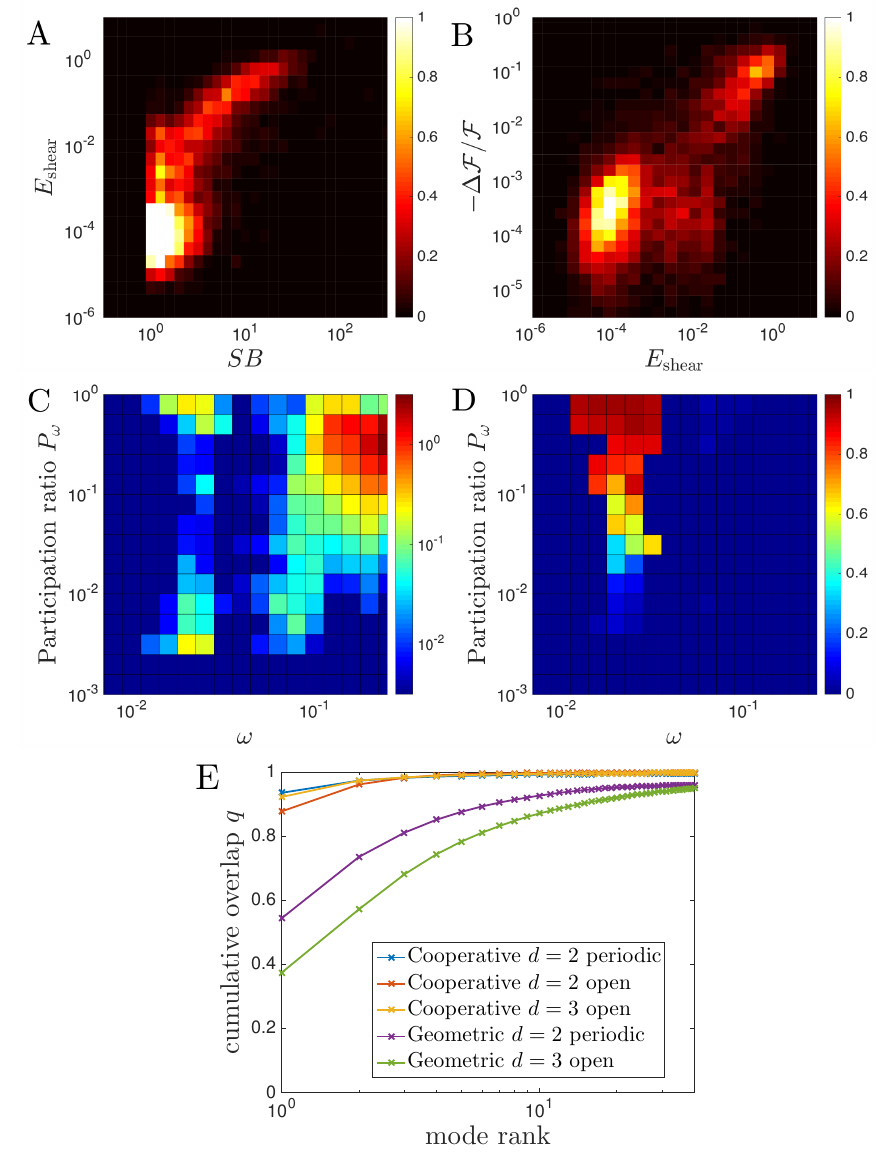} %
\caption{\small{Histogram of network nodes displaying (A) a given strain B-factor $SB$ and shear intensity $E_{\rm shear}$ (showing that most of the strain induced by the response to binding occurs in regions where the material is soft) and  (B) a given shear intensity $E_{\rm shear}$ and normalized fitness cost $-\Delta{\cal F}/F$ (showing that mutations are costly where the response strain is localized). The color bar indicates the relative abundance of the data points. (C) Density of vibrational modes $D(\omega,P_\omega)$ and (D) the  overlap $q(\omega,P_\omega)$ as a function of both frequency $\omega$ and participation ratio $P_\omega$, revealing the presence of a soft extended mode on which most of the response projects to.
For (A,B,C,D), the statistics is done over all 25 families of solutions found in the cooperative task in two dimension with a periodic boundary. 
(E) Cumulative overlap on the first $r$ modes with strongest overlap, where $r$ is denoted the rank. Results are shown both for the cooperative and the geometric tasks, for all dimensions and boundary conditions. 
}}\label{universal}
\end{figure}

{\bf Universal features of cooperative designs: } {Our in-silico evolution scheme generates different designs, as illustrated with the examples above. However, all these designed architectures follow the same principles, which we list in the following. These principles are systematically tested by averaging on the 25 families found in two-dimensions with periodic boundaries in Fig.\ref{universal}. The same analysis holds for other boundary conditions and in three dimensions as well, as documented in Supplemental Material Section E:}

\begin{itemize}
\item The system separates into a rigid and a soft manifold, {as observed in a class of proteins \cite{Tama00} and in protein models \cite{McLeish13}}.  


\item The  strain associated with the allosteric response is small in the rigid manifold (indicating rigid body or long-wavelength motion), while it is large in the soft manifold. {Both properties are apparent in Fig.~\ref{universal}A,  showing the two-dimensional density of nodes found with a given strain B-factor (reflecting the local rigidity) and strain intensity (reflecting the strain induced by the allosteric response).  This histogram displays a branch of soft nodes, where the strain B-factor is large and positively correlated to the strain intensity. }


\item In all cases, {the mutation cost is high precisely in these locations where the system is soft and where the strain intensity is large, as illustrated in Fig.~\ref{universal}B.} 

\item Most importantly, the daily-life examples we provided all have a common point: they display a single {\it mechanism}, i.e. a very soft elastic mode. We observe that this is also true in our cooperative architectures: there is always a single {\it soft and extended} mode along which most of the response projects to. 
This fact is already apparent in the decomposition of the response on vibrational modes shown in Figs.~\ref{cooppb}H,   Fig.~\ref{coopob}H and Fig.~\ref{coop3D}K. 
{It is studied systematically in Fig.~\ref{universal}C and Fig.~\ref{universal}D showing respectively the density of vibrational modes $D(\omega,P_\omega)$ and the  overlap $q(\omega,P_\omega)$ as a function of both frequency $\omega$ and participation ratio $P_\omega$. Fig.~\ref{universal}C shows a peak of extended (large $P_\omega$) modes at low $\omega$, Fig.~\ref{universal}D shows that most of the response projects precisely on these modes. We find that essentially one mode governs the response. This result can also be visualized by classifying modes for each system by decreasing  overlap $q$, and by representing the cumulative overlap (the sum of $q_\omega$ for the $r$ modes with the largest overlap) as a function of the rank $r$, as illustrated in  Fig.~\ref{universal}E. In average, the first mode captures more than 90\% of the response. }
\end{itemize}

{It is interesting to note that many properties of materials optimized to be cooperative, whose specific property is to display a single soft elastic mode controlling function,  differ from materials studied previously optimized to propagate a given strain~---~below we will refer to both cases as ``cooperative''  and ``geometric'' designs. An extensive comparison is performed in Supplemental Material Section F and G.  Salient differences include that: ({a}) the magnitude of the response is essentially constant in space in cooperative designs (it decays by five fold or more in  geometric designs) ({b}) the cooperative design is symmetric: binding at the allosteric or at the active site leads to a very similar response (whereas elastic information cannot propagate from the active site to the allosteric site in  geometric designs) ({c})  the cooperative design responds much more specifically than the geometric ones (in the latter case, imposing a strain anywhere in the material typically lead to a strong displacement at the active site) and ({d}) for geometric designs, the response does not correspond to a single soft elastic mode, but to a few of them, as already apparent in Fig.~\ref{universal}E.}

%
%
%

To explain the universal features of cooperative designs, and to predict the frequency of the soft extended mode controlling the response, we now investigate the optimality of designs.

\section*{Theory}

{\bf Absence of design:} We now argue that in a continuous elastic medium~---~where no design is involved~---~cooperativity decreases very rapidly with the distance $L$ between the allosteric and active sites. Any imposed local strain can be decomposed into multipole moments (dipole and higher), and the slower decaying response in the far field{~---~sufficiently distant from the source~---~}is dipolar{, since  higher multipoles decay faster}. {To model the perturbation induced by ligand binding, }we may thus consider without loss of generality two dipoles each of magnitude $f c$, where $f$ is the applied force and $c$ the distance over which these are exerted. 
Here we  give a simple scaling argument for a medium with elastic modulus $G$. As mentioned earlier, for $L\gg c$ we have $E_{\rm coop}\sim \langle dR|F\rangle$ where $|dR\rangle$ is now the dipolar response induced by the first dipole, of magnitude $dR(r)\sim \frac{fc}{r^{d-1} G}$, and $|F\rangle$ the force field of the second dipole. Since $|F\rangle$ is dipolar its scalar product on $|dR\rangle$ acts as a derivative taken at $r=L$, and one obtains $E_{\rm coop}\sim \frac{f^2c^2}{GL^d}\sim L^{-d}$, i.e. a very rapid decay with distance.  {This result is confirmed numerically for the case of a crystalline network in the Section D of Supplemental Material.}

\begin{figure}[htbp]
\includegraphics[width=\columnwidth]{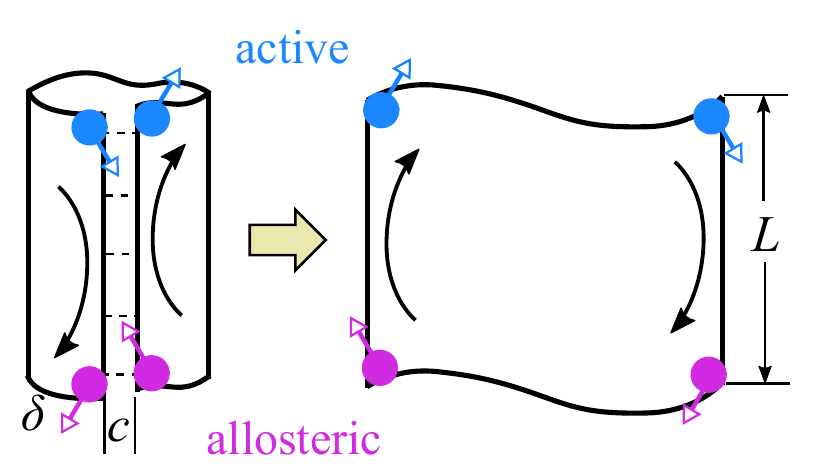}
\caption{\small{ In a cylindrical geometry, a mechanism~---~or zero mode~---~can be constructed by slicing the cylinder, as can be achieved by creating a cut of length $L$ and width $c$. One then obtains an object with the topology of a square, which now displays an additional  zero mode corresponding to a rigid rotation.   If the cut is filled up with a soft elastic material, the mode gets a finite frequency. As long as it is small, imposing a local displacement as indicated in the figure at the allosteric or at the active site will be dominated by this mode and will lead to essentially the same response. Thus $E^{{\cal A}c}\approx E^{{\cal A}l}\approx E^{{\cal A}c, {\cal A}l}$ and $E_{\rm coop}\approx E^{{\cal A}l}$.}
}\label{cutcylinder}
\end{figure}

\begin{figure}[htbp]
\includegraphics[width=1.\columnwidth]{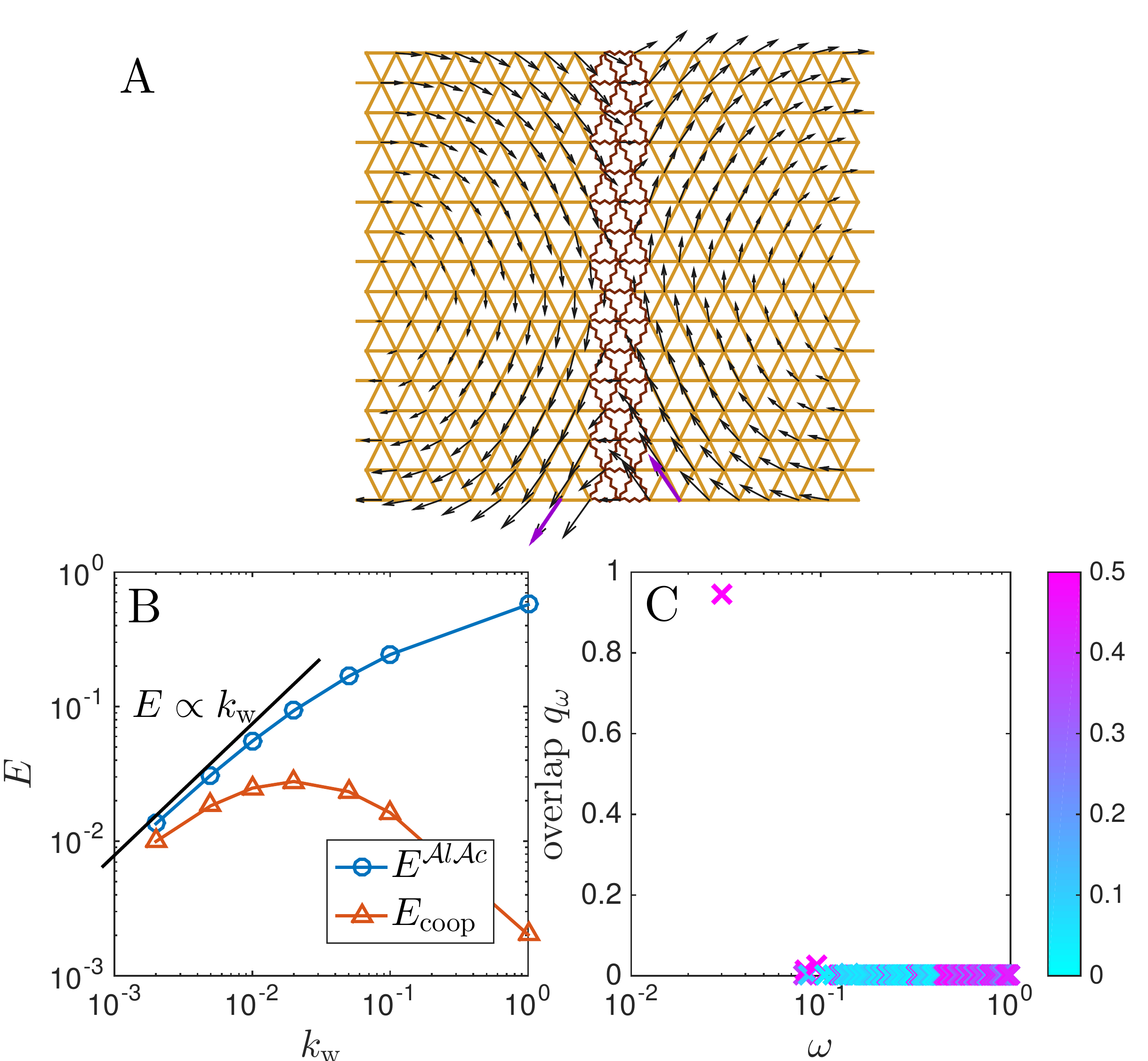}
\caption{\small{(A) We build a shear architecture using a triangular lattice with a soft band, where the springs have a stiffness $k_{\rm w}\ll k=1$ { such that the network modulus is proportional to the spring stiffness $G_{\rm w}/G=k_{\rm w}/k$}. The imposed displacement at the allosteric site is shown in purple arrows, and the associated response in black.
Here $k_{\rm w}=0.05$, $L=16$ and  $c=L/10$. (B) Energy of simultaneous binding $E^{{\cal A}c, {\cal A}l}$ and cooperative energy $E_{\rm coop}$ versus $k_{\rm w}$ for $L=32$. We confirm that $E_{\rm coop}$ depends non-monotonically on $k_{\rm w}$. 
(C) Overlap between the response and the eigenmodes $q_\omega$ {\it vs} mode frequency {${\omega}$ at optimal $k_{\rm w}^*=0.036$ for $L=32$ and $c=L/10$}, colored as a function of their participation ratio $P_\omega$. 
}
}\label{shear}
\end{figure}

{\bf Illustration of optimal cooperativity: Shear architecture. } We now show that cooperativity can be greatly improved if the material presents a very soft extended mode. For illustration we consider the  geometry of Fig.\ref{cutcylinder} where a cylinder of elastic modulus $G$ is cut on its length $L$, by a band of width $c$. This generates a zero mode corresponding to the rotation of a square. If displacements at the active or allosteric sites {of size $\delta$} are imposed as illustrated in Fig.\ref{cutcylinder}, they will only couple to that mode (since it costs no energy), and lead to the same response. This statement will be true even if the band of width $c$  is filled up with soft material of elastic modulus $G_{\rm w}$, as long as it is small enough (see below). Thus we have $E^{{\cal A}c}\approx E^{{\cal A}l}\approx E^{{\cal A}c, {\cal A}l}$ implying $E_{\rm coop}\approx E^{{\cal A}l}$, which can be readily estimated as the amount of elastic energy stored in the soft band, i.e. $E_{\rm coop}\sim \frac{L}{c}G_{\rm w}\delta^2$.

 This results implies that $E_{\rm coop}=0$ when the material presents a mechanism (i.e. $G_{\rm w}=0$), but increases with $G_{\rm w}$. This argument eventually breaks down, however, when it becomes more favorable to deform the rigid material and to couple to other modes in the system. This takes place when the energy of deforming a continuous medium of modulus $G$, $E_{cont}\sim G\delta ^2/\ln(L/c)$ becomes smaller than the energy associated with the soft mode $ \frac{L}{c}G_{\rm w}\delta^2$. Comparing these two expressions we get a cross-over for $G_{\rm w}=G_{\rm w}^*$ with:
 \be
 G_{\rm w}^*\sim \frac{c G}{L \ln(L/c)}
 \ee
 For $G_{\rm w}\gg G_{\rm w}^*$, the role of the soft mode become negligible and the system will respond as a  {homogeneous} elastic material (whose cooperativity is small as described above).
 Thus cooperativity will be maximal for $G_{\rm w}\approx G_{\rm w}^*$, leading to an optimal cooperativity of order:
 \be
 E^*_{\rm coop}\sim \frac{G\delta^2}{\ln(L/c)}
 \label{eq_eopt}
 \ee
 {This result is confirmed numerically in the Section D of Supplemental Material.}
 
 { The small energy of the response to binding  for large $L$ described by Eq.\ref{eq_eopt} implies the presence of a soft elastic mode, which is relevant experimentally. It can be detected in the vibrational spectrum of the protein, and implies large thermal fluctuations. Such fluctuations, in a harmonic approximation, are inversely proportional to the corresponding eigenvalue of the stiffness matrix, of order $\lambda^*\approx  E^*_{\rm coop}/ ||\delta R||^2$
 where $||\delta R||^2$ is the square norm of the allosteric response.  For the shear mode considered $||\delta R||^2\sim L^2 \delta^2$ since all particles are moving by a distance of order $\delta$, leading to $\lambda^*\sim 1/(L^2\ln(L/c))$.   For the vibrational spectrum such a small eigenvalue will lead to a low frequency $\omega^*$. Assuming for simplicity that all the particles have identical mass leads to:}
 {
 \be
 \omega^*\sim \sqrt{\lambda^*/m}\sim\frac{1}{L\ln^{1/2}(L/c)}
 \ee}
 which is thus much softer that the lowest-frequency plane wave modes, of frequency $1/L$ \footnote[1]{We expect this result to hold both for  the vibrational spectrum of the protein, or for that of the protein-ligand complex, for which  the frequency of the soft mode is  higher but of similar magnitude in our framework.}.
 
It is straightforward to extend these results to three dimensions in the geometry of a shear plane, where we find  $E^*_{\rm coop}\sim Gc\delta^2 $ which does not decay with distance,
and $\omega^*\sim L^{-3/2}$ which is now even much smaller than plane waves modes, thus justifying why the spectrum of our materials show an isolated soft extend mode at low frequency.  These results are tested in Fig.~\ref{shear} for $d=2$, which confirms that cooperativity is optimal for a finite frequency of the soft extended mode.

{\bf Principles of optimal cooperativity:}  
Overall, the common principle  emerging from this study is that optimal cooperativity results from the following antagonist effects. On the one hand, the architectures are such that they nearly present an extended mechanism. Because this mode is much softer than others, an imposed strain strongly couples to it, thus allowing to transfer the elastic information over long distances. On the other hand, if this extended soft mode is too soft, the elastic costs associated with binding become too small, leading to a small cooperative energy. As a result, there is an optimal frequency scale for cooperativity.

This idea leads to a natural explanation for the empirical facts listed in the introduction. Indeed shear and hinge designs (i) are clear realizations of this principle, which implies the presence of a soft extended modes at low frequency, consistent with observation (ii).

{We expect that our main result, i.e. the existence of an optimal vibrational frequency for cooperativity, will hold true when non-linearities are taken into account. This prediction can be tested using a combination of
molecular dynamics (MD) and experiments. MD can be used to measure the effect of point mutations on the thermal fluctuations along the relevant normal mode, and experiments can measure the effect of the same mutation on cooperativity.  In the spirit of Fig.\ref{shear}.B, we predict that there is an optimal magnitude of fluctuations for cooperativity to function properly. It would be very interesting to test if proteins function close to this optimum.}

{\bf Fluctuation-driven cooperativity:}  Finally, as pointed out in {\cite{Cooper84,McLeish13}}, the existence of a soft extended modes of frequency $\omega^*$ leads to the possibility of a cooperative effect with no mean displacement at play, once thermal effects are accounted for (iii). Indeed binding at the active site will hinder motion and increase the soft mode frequency, leading to an entropic cost that can be diminished if binding already took place at the allosteric site. Let us define $\omega_{{\cal A}l}$, $\omega_{{\cal A} c}$ and $\omega_{{\cal A}c, {\cal A}l}$ the frequencies of the soft mode after binding at the allosteric site, active site and both respectively. We can estimate these quantities as $\omega_{{\cal A}l}^2= \omega^*{}^2+e_{{\cal A}l}$,  $\omega_{{\cal A}c}^2= \omega^*{}^2+e_{{\cal A}c}$ and $\omega_{{\cal A}c, {\cal A}l}^2= \omega^*{}^2+e_{{\cal A}l}+e_{{\cal A}c}$ where $e_{{\cal A}l}$ ($e_{{\cal A}c}$) characterizes the additional energy  required for the mode to move when a ligand is bound at the allosteric (active) site. Assuming harmonic dynamics, the entropy of a normal mode of frequency $\omega$ reads $S=k_B \ln(k_BT/\hbar\omega)$.  Using this expression, one can now estimate the cooperative free energy $\Delta\Delta F=-T\Delta\Delta S=k_BT \ln(\omega_{{\cal A}c, {\cal A}l}\omega^*/\omega_{{\cal A}l}\omega_{{\cal A}c})=-k_BT \ln(1- e_{{\cal A}c}e_{{\cal A}l}/(\omega^*{}^2+e_{{\cal A}l})( \omega^*{}^2+e_{{\cal A}c}))$ which can indeed be large if $\omega^*{}^2$ is small compared to both $e_{{\cal A}l}$ and $e_{{\cal A}c}$.

\section*{Conclusion and outlook}

We have used in-silico evolution to design materials which are highly cooperative. Strikingly, the architectures found differ greatly from materials optimized to propagate a geometrical information over long distances. {The latter architectures are based on the emergence of a lever that amplifies the mechanical signal where it is desired, which may be relevant in proteins whose task is to trigger large motions when a ligand binds~---~e.g. to close an ion channel. By contrast, we predict that proteins optimized to be cooperative should display different architectures, including shear and hinge designs which are well-known in the literature. }Intriguingly, we find that there is a great variety of possible functioning architectures, especially in the three dimensional case. However, they all function along the same principle: they nearly display an extended mechanism, whose frequency should be neither too large nor too small for optimal function {to occurr}.

Our approach rationalizes several empirical observations on allosteric proteins and it also makes testable predictions. In particular, we predict that a single soft extended mode contributes to function, {whose frequency should decrease with protein size.} We find that this prediction is hard to test stringently from a spectral decomposition of the allosteric response alone, because localized soft modes (typically near the surface of the system) can hybridize with the relevant mode if they lie at similar frequencies. As a result, the response appears to project on a few modes (despite the localized modes being irrelevant for function) instead of one. 

Recent methods have been developed in computer science  to clean-up spectra of localized modes~---~see e.g.~\cite{ZhangPan16} in the field of community detection. An exciting path forward is to adapt these methods to proteins, allowing one to test if a single extended mode indeed contributes to allostery. Ultimately, this suggests a mechanical approach to discover de novo allosteric proteins, as those in which a single extended mode lies at low frequency in the cleaned-up spectrum. Such an analysis would further predicts where  mutations would affect function: we have observed that most damaging mutations hinder the allosteric response, and take place where the extended mode generates high shear.

\section*{Supporting Citations}
Reference~\cite{Landau86} appears in the Supporting Material.

\section*{Author Contributions}

L.Y. and M.W. conceived the project and M.W.  supervised  research. L.Y., R.R. and M.W. developed the theory. L.Y., R.R. and C.B. performed the computations and developed  the numerical methods. All authors analyzed the results and wrote the final manuscript. 
L.Y. and R.R. contributed equally to this work. 

\section*{Acknowledgments}

We thank J.P. Bouchaud, B. Bravi, S. Cocco,  T. De Geus,  P. De Los Rios, S. Flatt, W. Jie, D. Malinverni, R. Monasson,  M. Popovi\'c, S. Zamuner, Y. Zheng for discussions. L.Y. is supported by the Gordon and Betty Moore Foundation under Grant No. GBMF2919 and in part by the National Science Foundation under Grant No. NSF PHY-1748958.  M.W. thanks the Swiss National Science Foundation for support under Grant No. 200021-165509 and the Simons Foundation Grant ($\#$454953 Matthieu Wyart). This material is based upon work performed using computational resources supported by the ``Center for Scientific Computing at UCSB'' and NSF Grant CNS-0960316, and by the ``High Performance Computing at NYU''.

\bibliography{abbreviated}

\begin{thebibliography}{38}
\providecommand{\url}[1]{\texttt{#1}}
\providecommand{\urlprefix}{ }

\bibitem[Monod et~al.(1965)Monod, Wyman, and Changeux]{Monod65}
Monod, J., J.~Wyman, and J.-P. Changeux, 1965.
\newblock On the nature of allosteric transitions: a plausible model.
\newblock \emph{Journal of Mol. Biol.} 12:88--118.

\bibitem[Changeux and Edelstein(2005)]{Changeux05}
Changeux, J.-P., and S.~J. Edelstein, 2005.
\newblock Allosteric mechanisms of Signal Transduction.
\newblock \emph{Science} 308:1424--1428.

\bibitem[Amor et~al.(2016)Amor, Schaub, Yaliraki, and Barahona]{Amor16}
Amor, B.~R., M.~T. Schaub, S.~N. Yaliraki, and M.~Barahona, 2016.
\newblock Prediction of allosteric sites and mediating interactions through
  bond-to-bond propensities.
\newblock \emph{Nat. Commun.} 7.

\bibitem[Halabi et~al.(2009)Halabi, Rivoire, Leibler, and
  Ranganathan]{Halabi09}
Halabi, N., O.~Rivoire, S.~Leibler, and R.~Ranganathan, 2009.
\newblock Protein sectors: evolutionary units of three-dimensional structure.
\newblock \emph{Cell} 138:774--786.

\bibitem[Nussinov and Tsai(2013)]{Nussinov13}
Nussinov, R., and C.-J. Tsai, 2013.
\newblock Allostery in disease and in drug discovery.
\newblock \emph{Cell} 153:293--305.

\bibitem[Liang and Dill(2001)]{Liang01}
Liang, J., and K.~A. Dill, 2001.
\newblock Are proteins well-packed?
\newblock \emph{Biophys. J.} 81:751--766.

\bibitem[Gerstein et~al.(1994)Gerstein, Lesk, and Chothia]{Gerstein94}
Gerstein, M., A.~M. Lesk, and C.~Chothia, 1994.
\newblock Structural mechanisms for domain movements in proteins.
\newblock \emph{Biochemistry} 33:6739--6749.

\bibitem[Perutz(1970)]{Perutz70}
Perutz, M., 1970.
\newblock Stereochemistry of Cooperative Effects in Haemoglobin: Haem--Haem
  Interaction and the Problem of Allostery.
\newblock \emph{Nature} 228:726--734.

\bibitem[Xu et~al.(2003)Xu, Tobi, and Bahar]{Xu03}
Xu, C., D.~Tobi, and I.~Bahar, 2003.
\newblock Allosteric changes in protein structure computed by a simple
  mechanical model: hemoglobin T-R2 transition.
\newblock \emph{Journal of Mol. Biol.} 333:153--168.

\bibitem[Mitchell et~al.(2016)Mitchell, Tlusty, and Leibler]{Mitchell16}
Mitchell, M.~R., T.~Tlusty, and S.~Leibler, 2016.
\newblock Strain analysis of protein structures and low dimensionality of
  mechanical allosteric couplings.
\newblock \emph{Proc. Natl. Acad. Sci.} 201609462.

\bibitem[Goodey and Benkovic(2008)]{Goodey08}
Goodey, N.~M., and S.~J. Benkovic, 2008.
\newblock Allosteric regulation and catalysis emerge via a common route.
\newblock \emph{Nat Chem Biol} 4:474--482.

\bibitem[Gandhi et~al.(2008)Gandhi, Chen, Scott~Mathews, and Di~Cera]{Gandhi08}
Gandhi, P., Z.~Chen, F.~Scott~Mathews, and E.~Di~Cera, 2008.
\newblock Structural identification of the pathway of long-range communication
  in an allosteric enzyme.
\newblock \emph{Proc. Natl. Acad. Sci.} 105:1832--1837.

\bibitem[McLaughlin~Jr et~al.(2012)McLaughlin~Jr, Poelwijk, Raman, Gosal, and
  Ranganathan]{mclaughlin12}
McLaughlin~Jr, R.~N., F.~J. Poelwijk, A.~Raman, W.~S. Gosal, and
  R.~Ranganathan, 2012.
\newblock The spatial architecture of protein function and adaptation.
\newblock \emph{Nature} 491:138--142.

\bibitem[Atilgan et~al.(2001)Atilgan, Durell, Jernigan, Demirel, Keskin, and
  Bahar]{Atilgan01}
Atilgan, A., S.~Durell, R.~Jernigan, M.~Demirel, O.~Keskin, and I.~Bahar, 2001.
\newblock Anisotropy of fluctuation dynamics of proteins with an elastic
  network model.
\newblock \emph{Biophys. J.} 80:505--515.

\bibitem[De~Los~Rios et~al.(2005)De~Los~Rios, Cecconi, Pretre, Dietler,
  Michielin, Piazza, and Juanico]{Rios05}
De~Los~Rios, P., F.~Cecconi, A.~Pretre, G.~Dietler, O.~Michielin, F.~Piazza,
  and B.~Juanico, 2005.
\newblock Functional dynamics of PDZ binding domains: a normal-mode analysis.
\newblock \emph{Biophys. J.} 89:14--21.

\bibitem[Zheng et~al.(2006)Zheng, Brooks, and Thirumalai]{Zheng06}
Zheng, W., B.~R. Brooks, and D.~Thirumalai, 2006.
\newblock Low-frequency normal modes that describe allosteric transitions in
  biological nanomachines are robust to sequence variations.
\newblock \emph{Proc. Natl. Acad. Sci.} 103:7664--7669.

\bibitem[Popovych et~al.(2006)Popovych, Sun, Ebright, and
  Kalodimos]{Popovych06}
Popovych, N., S.~Sun, R.~H. Ebright, and C.~G. Kalodimos, 2006.
\newblock Dynamically driven protein allostery.
\newblock \emph{Nature Structural \& Mol. Biol.} 13:831--838.

\bibitem[Cooper and Dryden(1984)]{Cooper84}
Cooper, A., and D.~Dryden, 1984.
\newblock Allostery without conformational change.
\newblock \emph{Eur. Biophys. J.} 11:103--109.

\bibitem[Tsai et~al.(2008)Tsai, del Sol, and Nussinov]{Tsai08}
Tsai, C.-J., A.~del Sol, and R.~Nussinov, 2008.
\newblock Allostery: Absence of a Change in Shape Does Not Imply that Allostery
  Is Not at Play.
\newblock \emph{Journal of Mol. Biol.} 378:1--11.

\bibitem[McLeish et~al.(2013)McLeish, Rodgers, and Wilson]{McLeish13}
McLeish, T.~C., T.~Rodgers, and M.~R. Wilson, 2013.
\newblock Allostery without conformation change: modelling protein dynamics at
  multiple scales.
\newblock \emph{Phys. Biol.} 10:056004.

\bibitem[Hemery and Rivoire(2015)]{Hemery15}
Hemery, M., and O.~Rivoire, 2015.
\newblock Evolution of sparsity and modularity in a model of protein allostery.
\newblock \emph{Phys. Rev. E} 91:042704.

\bibitem[Tlusty et~al.(2017)Tlusty, Libchaber, and Eckmann]{Tlusty16}
Tlusty, T., A.~Libchaber, and J.-P. Eckmann, 2017.
\newblock Physical Model of the Genotype-to-Phenotype Map of Proteins.
\newblock \emph{Phys. Rev. X} 7:021037.

\bibitem[Yan et~al.(2017{\natexlab{a}})Yan, Ravasio, Brito, and Wyart]{Yan17}
Yan, L., R.~Ravasio, C.~Brito, and M.~Wyart, 2017.
\newblock Architecture and coevolution of allosteric materials.
\newblock \emph{Proc. Natl. Acad. Sci.} 114:2526--2531.

\bibitem[Rocks et~al.(2017)Rocks, Pashine, Bischofberger, Goodrich, Liu, and
  Nagel]{Rocks17}
Rocks, J.~W., N.~Pashine, I.~Bischofberger, C.~P. Goodrich, A.~J. Liu, and
  S.~R. Nagel, 2017.
\newblock Designing allostery-inspired response in mechanical networks.
\newblock \emph{Proc. Natl. Acad. Sci.} 114:2520--2525.

\bibitem[Flechsig(2017)]{Flechsig17}
Flechsig, H., 2017.
\newblock Design of Elastic Networks with Evolutionary Optimized Long-Range
  Communication as Mechanical Models of Allosteric Proteins.
\newblock \emph{Biophys. J.} 113:558 -- 571.

\bibitem[Sigmund and Maute(2013)]{Sigmund13}
Sigmund, O., and K.~Maute, 2013.
\newblock Topology optimization approaches.
\newblock \emph{Struct. Multidiscip. Optim.} 48:1031--1055.

\bibitem[Sigmund(1997)]{Sigmund97}
Sigmund, O., 1997.
\newblock On the Design of Compliant Mechanisms Using Topology Optimization.
\newblock \emph{Mechanics of Structures and Machines} 25:493--524.

\bibitem[Nishiwaki et~al.(1998)Nishiwaki, Frecker, Min, and
  Kikuchi]{Nishiwaki98}
Nishiwaki, S., M.~I. Frecker, S.~Min, and N.~Kikuchi, 1998.
\newblock Topology optimization of compliant mechanisms using the
  homogenization method.
\newblock \emph{International Journal for Numerical Methods in Engineering}
  42:535--559.

\bibitem[Yan et~al.(2017{\natexlab{b}})Yan, Bouchaud, and Wyart]{Yan17a}
Yan, L., J.-P. Bouchaud, and M.~Wyart, 2017.
\newblock Edge mode amplification in disordered elastic networks.
\newblock \emph{Soft Matter} 13:5795--5801.

\bibitem[Yan and Wyart(2014)]{Yan14}
Yan, L., and M.~Wyart, 2014.
\newblock Evolution of covalent networks under cooling: contrasting the
  rigidity window and jamming scenarios.
\newblock \emph{Phys. Rev. Lett.} 113:215504.

\bibitem[Yan and Wyart(2015)]{Yan15a}
Yan, L., and M.~Wyart, 2015.
\newblock Adaptive elastic networks as models of supercooled liquids.
\newblock \emph{Phys. Rev. E} 92:022310.

\bibitem[Maxwell(1864)]{Maxwell64}
Maxwell, J., 1864.
\newblock On the calculation of the equilibrium and stiffness of frames.
\newblock \emph{Philos. Mag.} 27:294--299.

\bibitem[Gullett et~al.(2007)Gullett, Horstemeyer, Baskes, and Fang]{Gullett07}
Gullett, P., M.~Horstemeyer, M.~Baskes, and H.~Fang, 2007.
\newblock A deformation gradient tensor and strain tensors for atomistic
  simulations.
\newblock \emph{Modell. Simul. Mater. Sci. Eng.} 16:015001.

\bibitem[Jacobs et~al.(2001)Jacobs, Rader, Kuhn, and Thorpe]{Jacobs01}
Jacobs, D.~J., A.~J. Rader, L.~A. Kuhn, and M.~F. Thorpe, 2001.
\newblock Protein flexibility predictions using graph theory.
\newblock \emph{Proteins: Struct., Funct., Bioinf.} 44:150--165.

\bibitem[Yang et~al.(2009)Yang, Song, and Jernigan]{Yang09}
Yang, L., G.~Song, and R.~L. Jernigan, 2009.
\newblock Protein elastic network models and the ranges of cooperativity.
\newblock \emph{Proc. Natl. Acad. Sci.} 106:12347--12352.

\bibitem[Tama et~al.(2000)Tama, Gadea, Marques, and Sanejouand]{Tama00}
Tama, F., F.~X. Gadea, O.~Marques, and Y.-H. Sanejouand, 2000.
\newblock Building-block approach for determining low-frequency normal modes of
  macromolecules.
\newblock \emph{Proteins: Struct., Funct., Bioinf.} 41:1--7.

\bibitem[Zhang(2016)]{ZhangPan16}
Zhang, P., 2016.
\newblock Robust Spectral Detection of Global Structures in the Data by
  Learning a Regularization.
\newblock \emph{In} D.~D. Lee, M.~Sugiyama, U.~V. Luxburg, I.~Guyon, and
  R.~Garnett, editors, Advances in Neural Information Processing Systems 29,
  Curran Associates, Inc., 541--549.

\bibitem[Landau and Lifshitz(1986)]{Landau86}
Landau, L., and E.~Lifshitz, 1986.
\newblock {Theory of Elasticity}, volume~7 of \emph{Course of Theoretical
  Physics}.
\newblock Pergamon Press, Oxford, U.K.

\end{thebibliography}


\section*{Supplementary Material}

\setcounter{equation}{0}
\setcounter{figure}{0}
\renewcommand{\theequation}{S\arabic{equation}}
\renewcommand{\thefigure}{S\arabic{figure}}

\subsection{A. Embedding lattice}
\begin{figure}[htbp]
\centering
\includegraphics[width=6.5cm]{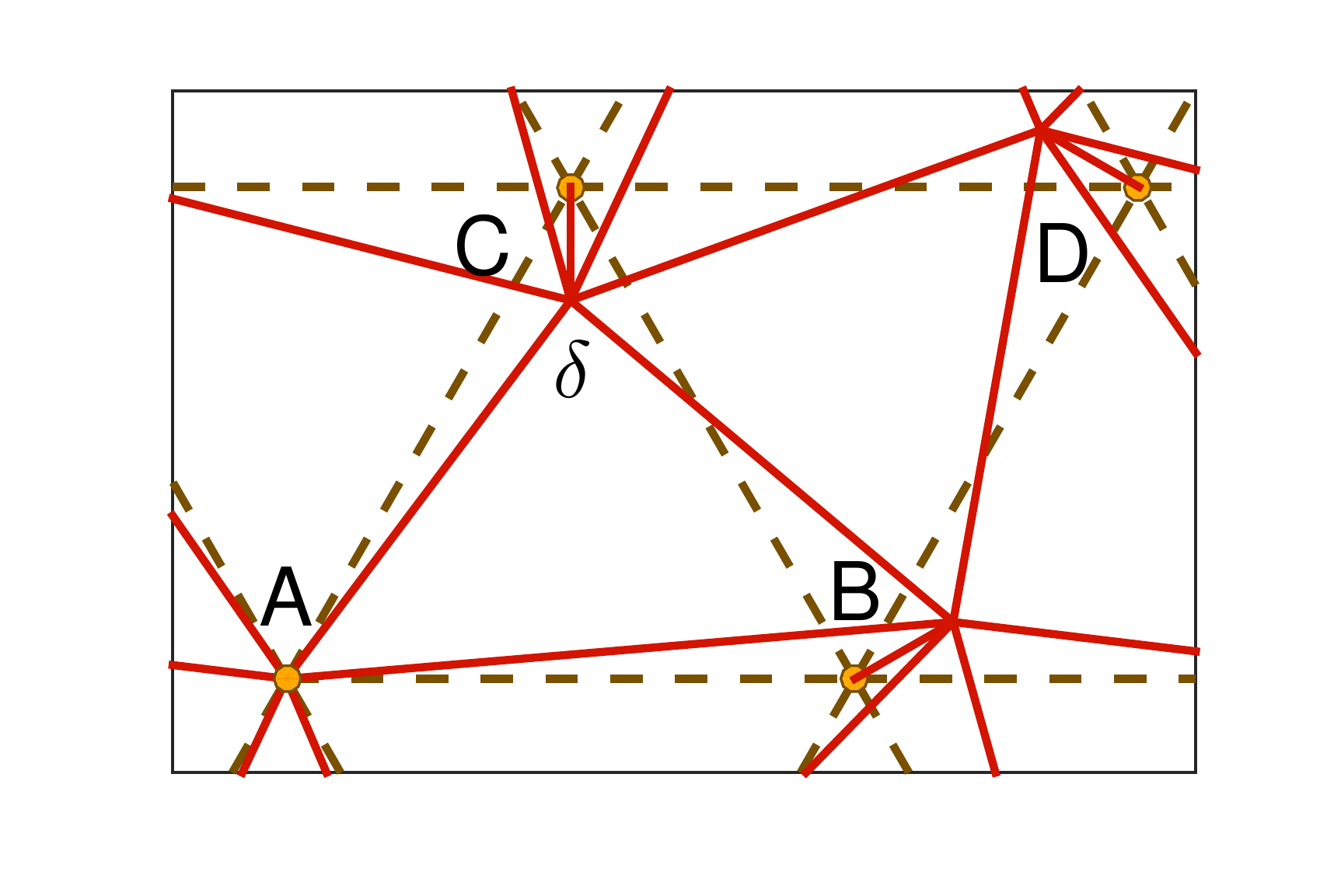}\\
\includegraphics[width=6cm]{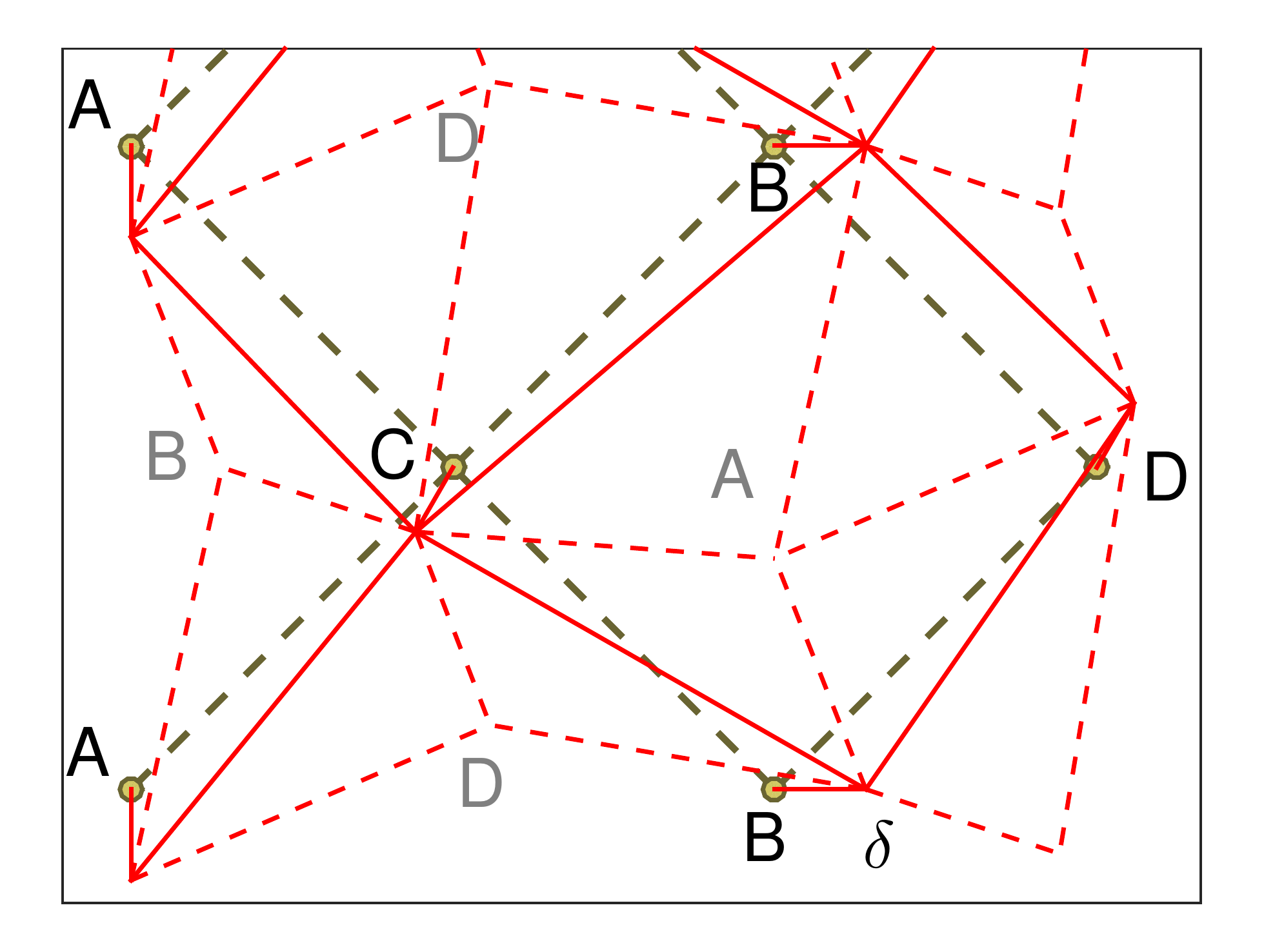}
\caption{Illustration of the distorted triangular lattice (top) and distorted FCC lattice (bottom). 
\label{distortion}}
\end{figure}

{\bf 2D triangular lattice.} 
In our model, we introduce a slight distortion of the lattice to remove long straight lines that occur in a triangular lattice. Such straight lines are singular and lead to unphysical localized floppy modes orthogonal to them.  
One can remove them by imposing a random displacement on the nodes. Instead, we distort the lines without introducing frozen disorder. We group nodes in lattice by four, labeled as A B C D in Fig.~\ref{distortion}. 
One group forms a cell of our distorted lattice. In each cell, node A stays in place,  while nodes B, C, and D move by some distance $\delta$: B along the direction perpendicular to BC, C along the direction perpendicular to CD, and D along the direction perpendicular to DB, as illustrated.  We set $\delta$ to  $0.2$, where the straight lines are maximally reduced with this distortion. 

{\bf 3D face-centered cubic (FCC) lattice.} 
We introduce a similar distortion to the FCC lattice. Again, we label the lattice nodes into four different types A B C D, as shown in one layer in the $z$ direction. The nodes are labeled in such a way that all 12 nearest neighbors of a node are different from it. For example, the center node in the bottom panel of  Fig.~\ref{distortion}, labeled as C, is connecting to two As and two Bs (in { solid lines}) in the layer and four Ds with two other As and Bs (in { dashed lines}) out of the layer. {To each layer in $z$ direction, there are two other layers, and in those two layers, D and A, B are located at the same $x$ and $y$. So we only see half of them (two D, one A and one B) connecting to C in a two dimensional projection along the $z$ direction in Fig.~\ref{distortion}.} We move all As along negative $y$ direction, all Bs along positive $x$ direction, all Cs along $(-\frac{1}{\sqrt{6}},-\frac{1}{\sqrt{2}},\frac{1}{\sqrt{3}})$, and Ds along $(\frac{1}{\sqrt{6}},\frac{1}{\sqrt{2}},-\frac{1}{\sqrt{3}})$ by $\delta=0.2$. As shown in Fig.~\ref{distortion}, all straight lines are thus perturbed without introducing quenched disorder. 

\subsection{B. Linear response of elastic networks}
{\bf Stiffness matrix.} 
Consider a displacement field $\delta\vec{R}_i\equiv\vec{R}_i-\vec{R}_{i0}$, where $\vec{R}_{i0}$ is the position of the node $i$ in the initial mechanical equilibrium. To the first order in $\delta\vec{R}_i$, the distance among neighboring nodes, defined as $r_{\langle ij\rangle}\equiv||\vec{R}_{i}-\vec{R}_{j}||$ between node $i$ and $j$, changes by 
\be
\delta r_{\langle ij\rangle} =r_{\langle ij\rangle}-r_{\langle ij\rangle,0}=\sum_{l}\ms_{\langle ij\rangle,l}\delta\vec{R}_l+o(\delta\vec{R}^2).
\label{dist}
\ee
$\ms_{\langle ij\rangle,\bullet}=\hat{n}_{\langle ij\rangle}(\langle i|-\langle j|)$, where $\hat{n}_{\langle ij\rangle}$ is the unit vector along link $\langle ij\rangle$ from $j$ to $i$, is the structure matrix. 

On the other hand, the force on a node is a composition of tensions,
\be
\vec{F}_i = \sum_j\hat{n}_{\langle ij\rangle}f_{\langle ij\rangle}=\sum_{\langle lm\rangle}\ms_{\langle lm\rangle,i}f_{\langle { lm}\rangle}.
\ee
For linear springs on the neighboring connections, $f_{\langle ij\rangle}=k_{\langle ij\rangle}\delta r_{\langle ij\rangle}$, the response force to the displacement is,
\be
|{\bf F}\rangle=\mm|\delta{\bf R}\rangle,
\label{fmr}
\ee
where the stiffness matrix $\mm_{i,j}=\sum_{\langle lm\rangle}k_{\langle lm\rangle}\ms_{\langle lm\rangle,i}\ms_{\langle lm\rangle,j}$, depends only on connection $|\sigma\rangle$ and the link directions. The elastic energy corresponds to the displacement field $|\delta{\bf R}\rangle$ of dimension $Nd$ is
\be
E = \frac{1}{2}\langle{\bf F}|\delta{\bf R}\rangle=\frac{1}{2}\langle\delta{\bf R}|\mm|\delta{\bf R}\rangle.
\ee

{\bf Linear response to an imposed displacement.}
When we impose a displacement on the subset ${\cal E}$ of { $N_{\cal E}$} nodes, $\delta{\bf R}^{\cal E}$, forces must be applied on these nodes. All other nodes adapt to a new mechanical equilibrium with no net forces on them, and follow a displacement $\delta{\bf R}_r$. Thus Eq.(\ref{fmr}) becomes, {for this choice of basis}:
\be
\lp\begin{array}{c}\vec{F}\\\vec{0}
\end{array}\rp
=\mm
\lp\begin{array}{c}
\delta{\bf R}^{\cal E}\\
\delta{\bf R}_r
\end{array}\rp.
\ee 
which leads to:
\be
\lp\begin{array}{c}
\vec{F}\\
\delta{\bf R}_r
\end{array}\rp
={\cal Q}^{-1}\mm
\lp\begin{array}{c}
\delta{\bf R}^{\cal E}\\
\vec{0}
\end{array}\rp
\label{qinv}
\ee
with 
\be
{\cal Q}_{ij} = \left\{
\begin{array}{lc}
\delta_{ij}& {\rm if}\ j\in{\cal E}\\
-\mm_{ij}& {\rm if}\ j\not\in{\cal{E}}
\end{array}\right..
\ee
{When {there} are floppy modes in the network, linear equation (\ref{qinv}) may not be solvable. In that case, ${\cal Q}^{-1}$ should be understood as the pseudo-inverse so that the network does not respond along the floppy directions (corresponding singular values are zero in ${\cal Q}$). Another possibility is to reduce singularity by imposing that each node also interacts with all its next nearest neighbors via weak springs of stiffness $k_{\rm w}\ll1$. Both methods lead to qualitatively identical results. For numerical costs, our results were computed using the second approach with $k_{\rm w}=10^{-4}$. So our stiffness matrix $\mm=\ms_\sigma^t\ms_\sigma+k_{\rm w}\ms_{\rm w}^t\ms_{\rm w}$.}

{\bf Regarding translations and rotations.}
{ When binding a ligand, the translational and rotational degrees of freedom (TR) of the nodes are not determined. 
If we write the TR degrees of freedom at the imposed nodes as $\Psi^{\cal E}$, a $d N_{\cal E}\times d_{TR}$ matrix, which is a set of vectors with $d_{TR}=6$ in $d=3$ and $d_{TR}=3$ in $d=2$, any imposed displacement giving the same shape change is then,
\be
\delta{\bf R}^{\cal E}=\delta{\bf R}^{\cal E}_0+\Psi^{\cal E}\cdot\vec{c}.
\ee
where $\delta{\bf R}^{\cal E}_0$ is purely determined by the shape change, $\delta{\bf R}^{\cal E}_0\cdot\Psi^{\cal E}=\vec{0}$,  and $\vec{c}$ is a parameter vector of dimension $d_{TR}$ to count TR contribution additional to the shape change. 
We can thus consider a new basis with a $dN_{\cal E}$ by $dN_{\cal E}$ transform matrix ${\cal U}$ to the original space on imposed nodes so that 
\be
\delta{\bf R}^{\cal E}={\cal U}\lp\begin{array}{c}\vec\delta_0\\\vec{c}\end{array}\rp,
\ee
translations and rotations are isolated from the shape change defined by $\vec{\delta}_0$.
 In this new basis, the forces $\vec{F}$ imposed on ${\cal E}$ obey total force and torque balance,
\be
{\cal U}^t\vec{F} = \lp\begin{array}{c}\vec{f}\\\vec{0}\end{array}\rp.
\ee
The linear response problem thus becomes, 
\be
\tilde{\cal Q}\lp\begin{array}{c}\vec{f}\\\vec{c}\\\delta{\bf R}_r\end{array}\rp = \lp\begin{array}{cc}{\cal U}^t&0\\0&\cal I\end{array}\rp{\cal M}\lp\begin{array}{c}\delta{\bf R}^{\cal E}_0\\\vec{0}\end{array}\rp
\ee
where 
\be
\tilde{\cal Q}_{ij}=\left\{\begin{array}{lc}
\delta_{ij} & {\rm if}\ j\in{\cal E}\setminus TR\\
-\tilde{\mm}_{ij} & {\rm otherwise}
\end{array}\right.
\ee}
with 
\be
\tilde{\mm} = \lp\begin{array}{cc}{\cal U}^t&0\\0&\cal I\end{array}\rp{\cal M}\lp\begin{array}{cc}{\cal U}&0\\0&{\cal I}\end{array}\rp.
\ee
{Note that given the separation of the two subspaces the matrix $\cal U$ is of dimension $d N_{\cal E}\times d N_{\cal E}$ and the matrix \cal I is $d(N - N_{\cal E})\times d(N -N_{\cal E})$, consistent with $\cal M$ being $dN \times dN$. }

\subsection{C. Computing the local strain tensor in a network}
In a continuous medium, a motion maps a point $\vec{X}$ in the reference configuration to a new point ${\vec{x}}$ in the current configuration, the strain tensor of the motion can thus be computed as,
\be
{\epsilon}_{ab}(\vec{X})=\frac{1}{2}\lp\frac{\partial\vec{x}}{\partial X_a}\cdot\frac{\partial\vec{x}}{\partial X_b}-\delta_{ab}\rp,
\ee
where $a$, $b$ labels the spatial dimension. 

In a discrete medium as networks, the problem is to compute the partial derivative $\overset\leftrightarrow{\Lambda}=\partial\vec{x}/\partial\vec{X}$ at node $i$ for especially non-lattice structures. Ideally, for any neighbor $j$ close enough in space,
\be
\Delta \vec{x}_{ij}=\overset\leftrightarrow{\Lambda}_i\cdot\Delta\vec{X}_{ij},
\label{eq_lamb}
\ee
where $\Delta\vec{X}_{ij}=\vec{R}_{i0}-\vec{R}_{j0}$ and $\Delta\vec{x}_{ij}=\vec{R}_i-\vec{R}_j$ in our model. 
We have $n_b$ number of such equations for $\overset\leftrightarrow{\Lambda}_i$ when $n_b$ neighbors are considered. 
So $\overset\leftrightarrow{\Lambda}_i$ are usually over-determined when we consider all nearest neighbors ($n_b=6$ for a $2\times2$ matrix in triangular lattice, and $n_b=12$ for a $3\times 3$ matrix in FCC lattice). 
Instead of solving Eq.(\ref{eq_lamb}), we define a mean squared error function \cite{Gullett07},
\be
MSE(i)=\sum_{j}(\Delta \vec{x}_{ij}-\overset\leftrightarrow{\Lambda}_i\cdot\Delta\vec{X}_{ij})^2w_j(i),
\ee
where we have kept a weight function $w_j(i)$ of node $j$ contribution to $i$ in general. Specifically, we set $w_j(i)=\frac{1}{n_b}$ for all nearest neighbors to $i$ on the original embedding lattice and $w_j=0$ otherwise. By minimizing the mean squared error with respect to $\overset\leftrightarrow{\Lambda}_i$, we have
\be
\overset\leftrightarrow{\Lambda}_i=\sum_j\Delta\vec{x}_{ij}\Delta\vec{X}_{ij}w_j(i)\cdot\lp\sum_j\Delta\vec{X}_{ij}\Delta\vec{X}_{ij}w_j(i)\rp^{-1},
\ee
and 
\be
\overset\leftrightarrow{\epsilon}(i)=\frac{1}{2}\lp\overset\leftrightarrow{\Lambda}_i^t\cdot\overset\leftrightarrow{\Lambda}_i-\overset\leftrightarrow{\delta}\rp.
\ee

{\subsection{D. Cooperative energy of two dipoles in a continuous elastic medium}
{\bf Elastic energy of a force monopole.} 
We make the simplifying assumption that the velocity field is divergence free (relaxing this assumption will not change the predicted scaling behaviors). The equation for the displacement field when a monopole force $\vec{f}$ is applied to a constant force over a spherical patch of radius $c$ then follows~\cite{Landau86}:
\be
\Delta\vec{u}=\nabla\cdot\nabla\vec{u} = -\frac{d}{G\Omega_dc^d}\vec{f},
\label{eq_monof}
\ee
where $G$ is the shear modulus, $\Omega_d$ is the solid angle of $d$ dimensional sphere. Defining $\vec{f}=f\hat{e}_y$, both force and displacement component are along $y$ direction, we then solve for the divergence of the displacement field using Gauss Theorem,
\be
\nabla u_y=\left\{\begin{array}{cc}-\frac{f}{G\Omega_dc^d}r\hat{e}_r, & r<c\\
-\frac{f}{G\Omega_d}\frac{1}{r^{d-1}}\hat{e}_r. &r\geq c
\end{array}\right.
\ee
The total energy of the monopole is approximately,
\be
E_m=G\int\rd^d\vec{r}(\nabla u_y)^2=\frac{f^2}{G\Omega_d}\left(\int_0^c\frac{r^{d+1}}{c^{2d}}+\int_c^R\frac{1}{r^{d-1}}\right)\rd r,
\label{eq_emono}
\ee
where $R$ defines the system size. In 2D, the integral is dominated by the second term,
\be
E_m=\frac{f^2}{2\pi G}\ln\frac{R}{c}.
\label{eq_em2}
\ee
In 3D and above, the integral of the second term converges in the large size limit $R\to\infty$, and it has the same scaling as the first term, 
\be
E_m=\frac{2df^2}{(d^2-4)\Omega_dG}c^{2-d}.
\label{eq_emd}
\ee
So the displacement $\delta$ can be achieved by an external force satisfying $\delta=\partial E_m/\partial f$, 
\be\ba
\delta&=f\frac{1}{\pi G}\ln\frac{R}{c};\qquad &d=2\\
\delta&=f\frac{4d}{(d^2-4)\Omega_dG}c^{2-d}.\qquad &d>2
\ea\ee}

{{\bf Elastic energy of a force dipole.}
To compare with the mechanism discussed in the main text, we hereby compute the cooperative energy of two dipoles of size $c$ separated by $L$ in a homogeneous medium, as illustrated in Fig.8 (main text).
Similar to the monopole energy computed above, we could define the dipole energy, 
\be
E_d=G\int\rd^d\vec{r}\frac{f^2}{\Omega_d^2G^2}\sum_{i=1}^d(x_{+,i}-x_{-,i})^2
=2E_m+\tilde{E}_d,
\label{eq_edipole}
\ee
where $x_+$, $x_-$, $y_+$, $y_-$ are components in $x$ and $y$ directions contributed by the $+$ monopole and $-$ monopole respectively in the dipole. So the dipole self-energy is,
\be
\tilde{E}_d=-\frac{2f^2}{\Omega_d^2G}\int\rd^d\vec{r}\sum_{i=1}^dx_{+,i}x_{-,i},
\label{eq_edip}
\ee
where the integral is over three regions, within $c$ to the $+$ monopole, within $c$ to the $-$ monopole, and the rest. One can show that the contributions of the first two regions inside monopoles scale as $c^{d+2}$, while the contribution of the remaining region scales as $c^{d+2}$, comparable. Outside of the monopoles, $x_{\bullet,i}\approx\frac{1}{r^d}x_i$, so 
\be
\tilde{E}_d\sim-\frac{f^2}{G}\int_c^{R}r^{d-1}\rd r\frac{1}{r^{2d-2}}\sim\left\{\begin{array}{cc}-\frac{f^2}{G}\ln\frac{R}{a} & d=2\\
-\frac{f^2}{G}c^{2-d} & d>2
\end{array}\right..
\label{eq_edi}
\ee}

{{\bf Cooperative energy of two force dipoles.} 
Similar to the way we computed the dipole self-energy Eq.(\ref{eq_edip}), the cooperative energy, which is defined as the extra energy from the interaction of two dipoles,  can be computed as
\begin{multline}
E_{\rm coop}=2E_d-E_{\rm tot}\\
=-\frac{2f^2}{\Omega_d^2 G}\int\rd^d\vec{r}\sum_{i=1}^d(x^0_{+,i}-x^0_{-,i})(x^L_{+,i}-x^L_{-,i}).
\label{eq_ecoop}
\end{multline}
where $x^L_{\bullet}$ are the contributions of the monopole at $L$. When $c/L\ll1$, the contribution outside of both the monopoles and the dipoles dominates the energy,
\begin{multline}
E_{\rm coop}\sim\frac{f^2}{G}\int_0^\infty\rd\rho\int_c^L\rd z\rho^{d-2}\frac{c^2}{[\rho^2+z^2]^{d/2}[\rho^2+(L-z)^2]^{d/2}}\\
\sim\frac{f^2c^2}{GL^d}\ln\frac{L}{c}.
\label{eq_ecoopL}
\end{multline}
}
{
For given displacement $\delta$ applied at the dipoles,
\be
E_{\rm coop}\sim\left\{\begin{array}{cc}G\frac{c^2\delta^2}{L^2\ln\frac{L}{c}} & d=2\\
G\frac{c^{2d-2}\delta^2}{L^d}\ln\frac{L}{c} & d>2
\end{array}\right.
\label{eq_ecoopL1}
\ee
showing that $E_{coop}$ decays as fast as $L^{-d}$ for two dipoles at a distance $L$ from each other (with weak logarithmic corrections in $d=2$).}

{{\bf Numerical verification.} 
\begin{figure}[htbp]
\includegraphics[width=\columnwidth]{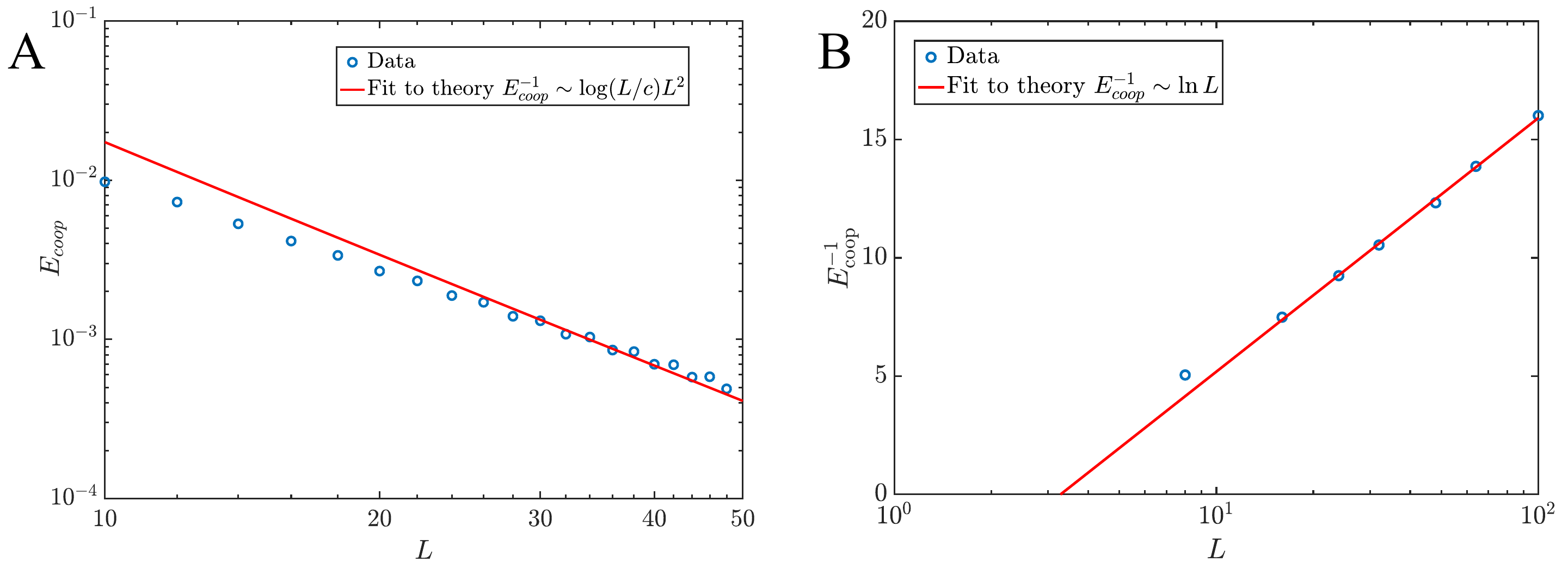}
\caption{\small{(A) Cooperative energy computed for a distorted crystal ($\delta=0.2$) of varying size $L$ with no mechanism. (B) Inverse of the cooperative energy for a crystal with a soft shear band presenting a mechanism, the softness of the band being chosen as the value of $k_w$ where the cooperative energy is optimal, see Fig. 9. The two different scalings predicted from continuous elastic media are fitted and shown as solid lines. }
}\label{ff0}
\end{figure}
In Fig.~\ref{ff0}(A)  we test our prediction for the cooperativity of a homogenous medium without any design (a distorted crystal with $\delta=0.2$), and confirm Eq.~\ref{eq_ecoopL1} for $d=2$. In Fig.~\ref{ff0}(B) we test our prediction for an optimal shear design, and confirm the very weal logarithmic decay of the cooperative energy in two dimensions, as described in Eq.11 in the main text. }

\subsection{E. Principles of cooperative designs: numerical tests}
\begin{figure}[htbp]
\includegraphics[width=\columnwidth]{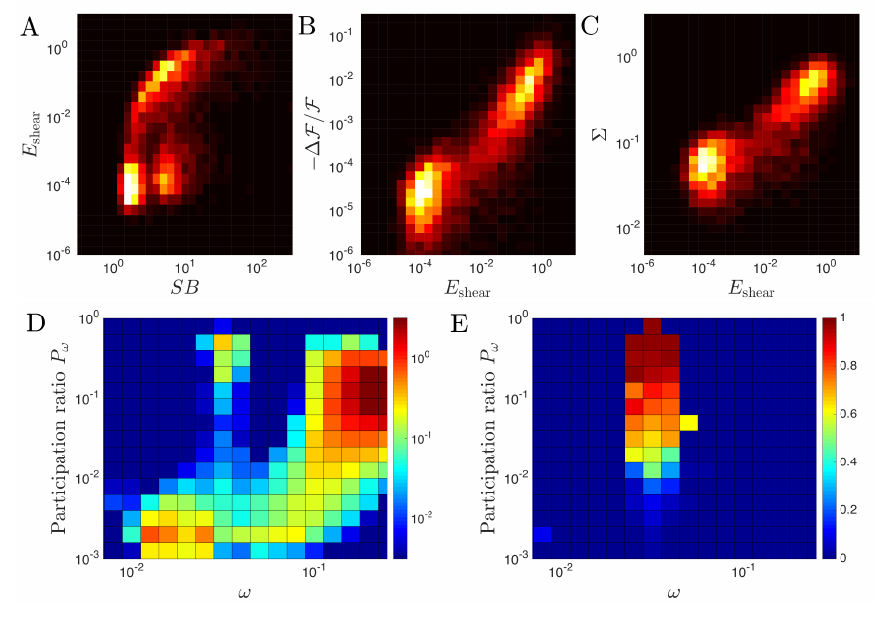}
\caption{\small{Same analysis as Fig.\ref{ff1} for $d=3$ and open boundaries. }
}\label{ff2}
\end{figure}

{In the main text we have listed the principles underlying the cooperative architectures, and tested them in Fig.7 in two dimensions, with a periodic boundary. The same results hold with open boundaries in $d=2$ (Fig.\ref{ff1}) and $d=3$ (Fig.\ref{ff2}).

\begin{figure}[htbp]
\includegraphics[width=\columnwidth]{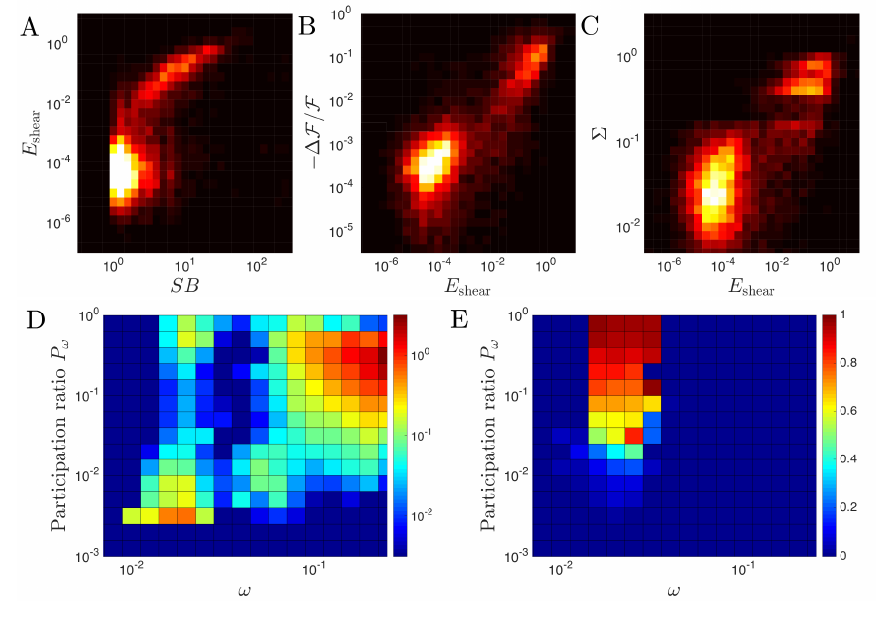}
\caption{\small{Analysis of the cooperative task in two dimensions with {open boundaries}:
histogram of network nodes displaying (A) a given strain B-factor $SB$ and shear intensity $E_{\rm shear}$ (showing that most of the strain induced by the response to binding occurs in regions where the material is soft), (B) a given shear intensity $E_{\rm shear}$ and normalized fitness cost $-\Delta{\cal F}/{\cal F}$ (showing that mutations are costly where the response strain is localized) and (C) a given  $E_{\rm shear}$ and conservation $\Sigma$ (showing that these same locations are highly conserved). (D) Density of vibrational modes $D(\omega,P_\omega)$ and (E)   overlap $q(\omega,P_\omega)$ as a function of both frequency $\omega$ and participation ratio $P_\omega$, revealing the presence of a soft extended mode on which most of the response projects to.}
}\label{ff1}
\end{figure}

\subsection{F. Geometric task}

\begin{figure*}[htbp]
\includegraphics[width=\linewidth]{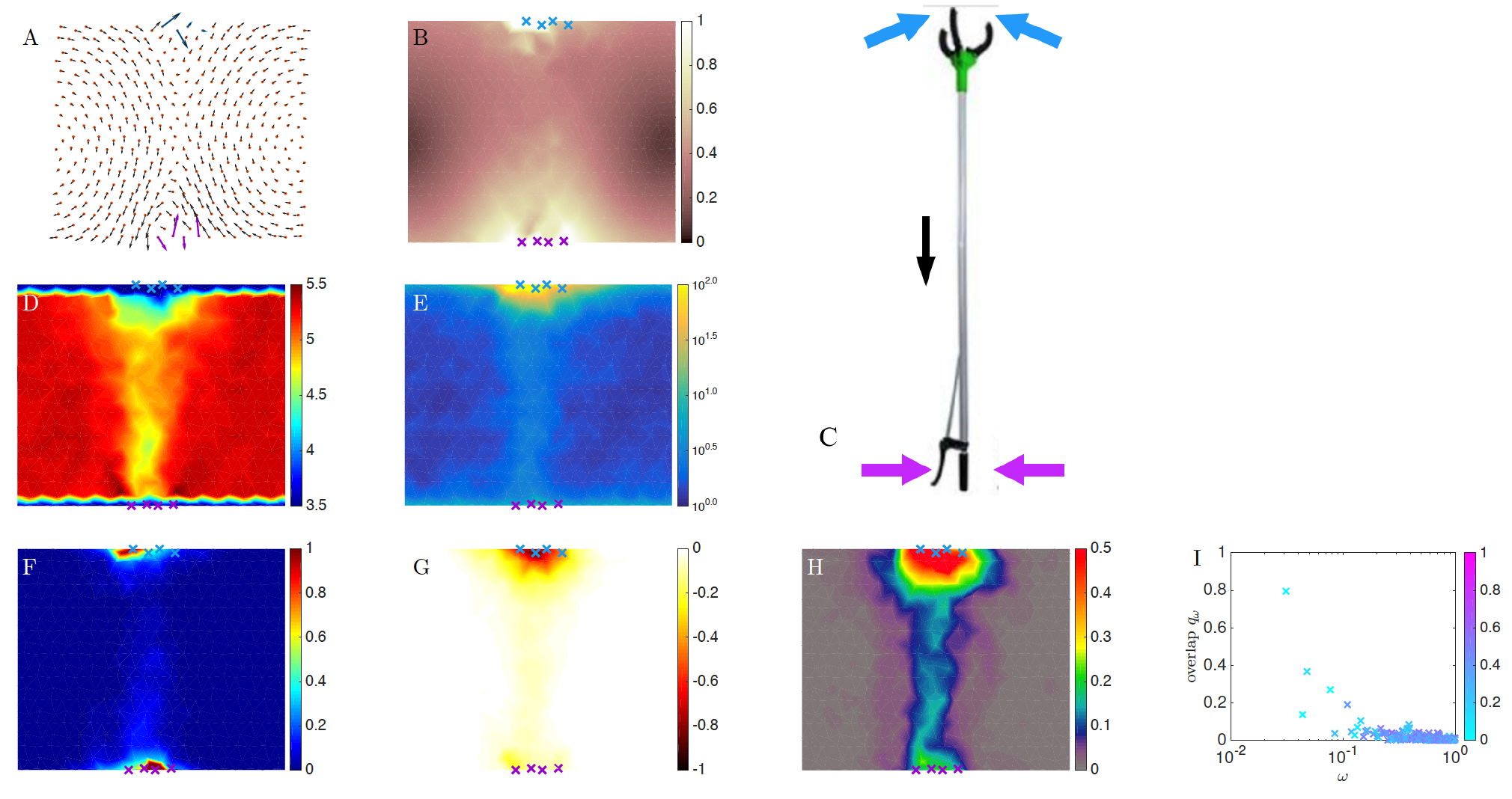}
\caption{\small{Two-dimensional geometric task with periodic boundaries. (A) The average response $\delta{\bf R}_r^{{\cal A}l}$ induced by binding at the allosteric site is shown in black arrows. (B) Map of the average magnitude of response $||\delta{\bf R}_r||$. (C) A fruit picker illustrates the combined mechanisms of edge mode lever and shear. (D) Map of the average coordination number $z$. (E) Map of the average strain B-factor $SB$. (F) Map of the average shear intensity $E_{\rm shear}$. (G) Map of the fitness cost of single site mutation normalized by the cost of random networks $\Delta F/F$. (H) Map of the average conservation $\Sigma$. (I) Decomposition  $q_\omega$ of the response on the vibrational modes $\omega$ in a specific solution, colored as a function of the participation ratio $P_\omega$.}
}\label{allos2D}
\end{figure*}

{Definition: networks perform the geometric task by minimizing a cost function that measures the deviation of the allosteric response $|\delta{\bf R}^{{\cal A}l}_r\rangle$ from a prescribed shape change located at the active site $|\delta{\bf R}^{{\cal A}c}\rangle$~\cite{Yan17},
\be
E(\sigma)\equiv\min_{\cal |{\bf U}\rangle}\sqrt{\sum_{i\in{{\cal A}c}}(\delta{\bf R}^{{\cal A}l}_{r,i}-\delta{\bf R}^{{\cal A}c}_i-{\cal {\bf U}}_i)^2},
\ee
where $|{\cal {\bf U}}\rangle$ is a global translation and rotation, which does not change the shape at the active site.  Here ${\cal A}c$ corresponds to four sites defining the active site.

We illustrate the method  by studying the case $d=2$ with periodic boundaries, as well as $d=3$ with free boundaries. 
Our results are averaged  over 25 runs with different initial conditions in $d=2$ and 10 runs in $d=3$.
}

\begin{figure*}[htbp]
\includegraphics[width=\linewidth]{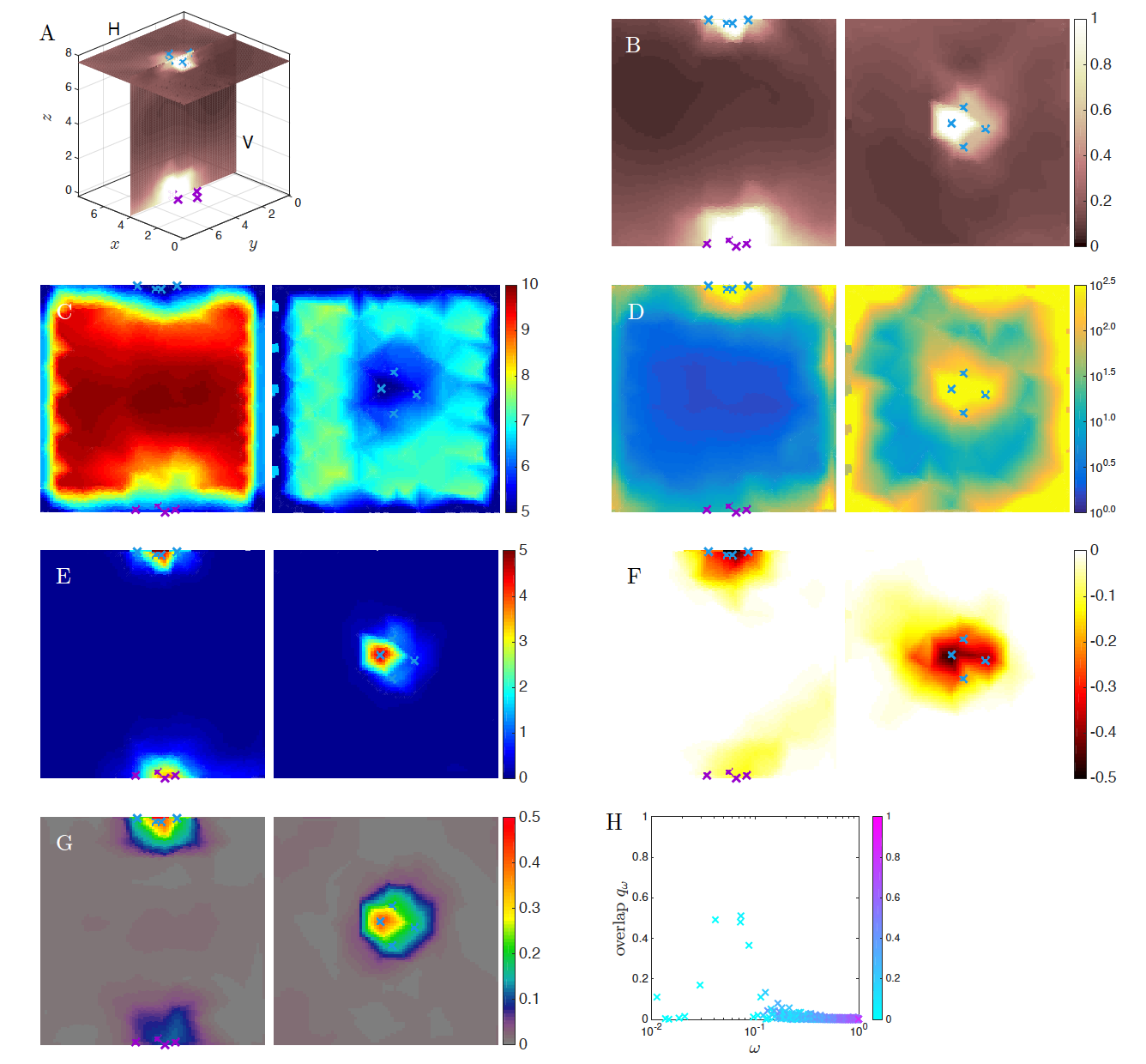}
\caption{\small{Three-dimensional geometric task with open boundaries. (A) Two dimensional sections of the 3D bulk: the vertical plane V and horizontal plane H are shown on left and right respectively in the following panels. (B) Map of the average magnitude of response $||\delta{\bf R}_r||$. (C) Map of the average coordination number $z$. (D) Map of the average strain B-factor $SB$. (E) Map of the average shear strain intensity $E_{\rm shear}$. (F) Map of the average fitness cost of single site mutation normalized by the cost of random networks $\Delta F/F$. (G) Map of the average conservation $\Sigma$. (H) Decomposition  $q_\omega$ of the response on the vibrational modes $\omega$ in a specific solution, colored as a function of the participation ratio $P_\omega$.}
}\label{allos3D}
\end{figure*}

{{\bf $d=2$:} This case is documented in Fig.~\ref{allos2D}, using the observables introduced in the main text.
The architecture presents  two main features. Most importantly, the response is non-monotonic and strongly amplified close to the active site, as shown in Fig.~\ref{allos2D}B. As demonstrated in \cite{Yan17a,Yan17}, this effect is induced by the presence of a marginally connected region with $z=4$  (Fig.~\ref{allos2D}D) which is thus very soft (Fig.~\ref{allos2D}E). It can be shown to act as a powerful lever \cite{Yan17a}. It is also highly conserved and leads to high mutation costs (Fig.~\ref{allos2D}G,H). 

Another aspect of the observed design is the emergence of a shear mode close to the allosteric site, as can be seen from the mean response (Fig.~\ref{allos2D}A) and from the map of the shear intensity (Fig.~\ref{allos2D}F). This response is caused by 
the mergence of a weakly-coordinated band above the allosteric site  (Fig.~\ref{allos2D}D).

Overall, the design is thus similar to that of a fruit-picker (Fig.~\ref{allos2D}C) where a stimulus (violet arrows) leads to a shear (black arrows) that couples to a head (black arrows), which acts as a lever and leads to a specific desired response.  This design leads to a more complex spectral signature where several modes typically contribute to the response (Fig.~\ref{allos2D}I), instead of one as for the cooperative designs discussed in the main text.}

{{\bf $d=3$:} 
The arguably most relevant case corresponds to $d=3$ with open boundaries, and is illustrated in  Fig.~\ref{allos3D}.
 As shown in Fig.~\ref{allos3D}A, we study the response by focusing on  two  sections: a vertical plane passing through both the allosteric and active sites, and a horizontal plane containing the active site. Once again, we find that the central aspect of the design is the emergence of a weakly-connected region with $z\approx 6$ (the isostatic value) surrounding the active site (Fig.~\ref{allos3D}C), and leading  to a very pronounced amplification of the response (Fig.~\ref{allos3D}B). This lever region is soft (Fig.~\ref{allos3D}D) and conserved (Fig.~\ref{allos3D}E,G). In that case, there is no evidence in the shear map of a hinge or shear motion in the material bulk (Fig.~\ref{allos3D}E). Fig.~\ref{allos3D}H shows that the lever design alone comes with a rather complex spectral decomposition of the response in which several modes contribute.}

\begin{figure*}[htbp]
\includegraphics[width=0.8\linewidth]{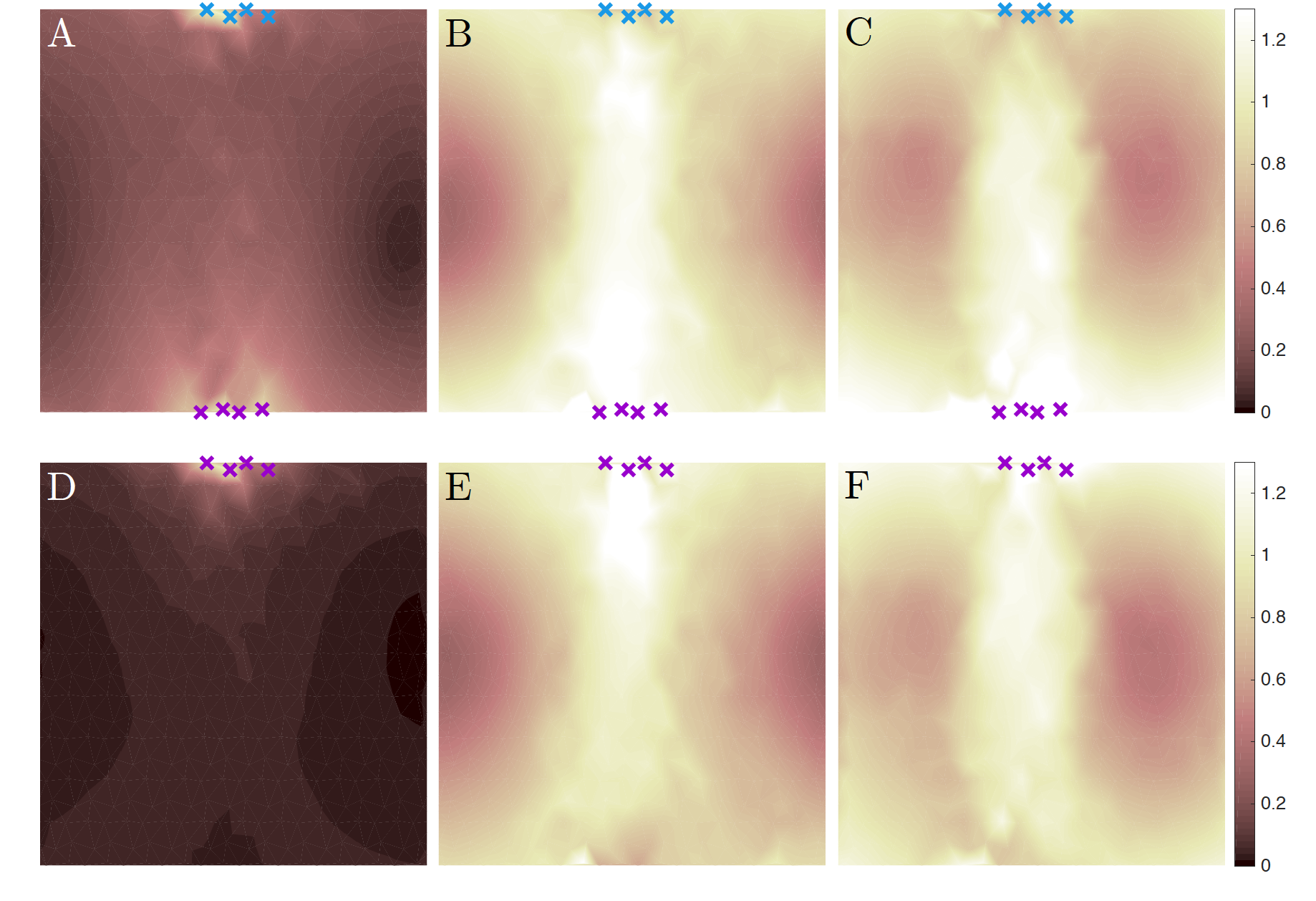}
\caption{\small{Magnitude of the response averaged over (A,D) geometric solutions {in periodic boundary}, (B,E) cooperative solutions in periodic boundary and (C,F) in open boundary. The stimulus strain is imposed at the nodes shown as the purple crosses in (A-C)- precisely where these materials have evolved to respond. By contrast, in  (D-F) the strain is imposed at the active site (purple crosses) where the material has not learnt to respond. For the geometric design, the response dies out very rapidly within the material, but for the cooperative design, the response is similar that obtained by stimulating the allosteric site shown above.}
}\label{monoasym}
\end{figure*}

\subsection{G. Comparison between cooperative and geometric designs}

{The key difference between the geometric  and the cooperative designs is that the former develops a lever, while the latter doesn't.  This fact leads to vastly different properties of the response to binding.

 {\bf Amplification of the response:} The map of the average magnitude of the response  shows a non-monotonic behavior between the allosteric and the  active sites (Fig.~\ref{monoasym}A) for geometric designs, not apparent for cooperative designs (as shown in Fig.~\ref{monoasym}B,C for both periodic and open boundaries respectively). 

 {\bf Symmetry  of the response:}  For cooperative designs, the response to binding at the active site is very similar to binding at the allosteric site (Fig.~\ref{monoasym}E,F), because both type of stimuli mostly couple to the single soft elastic mode in the system.  For the geometric design, this is not true at all: stimulating the material in the active site where it is soft has essentially no effect in the rest of the material, as shown in Fig.~\ref{monoasym}D. 
}

\begin{figure*}[htbp]
\includegraphics[width=\linewidth]{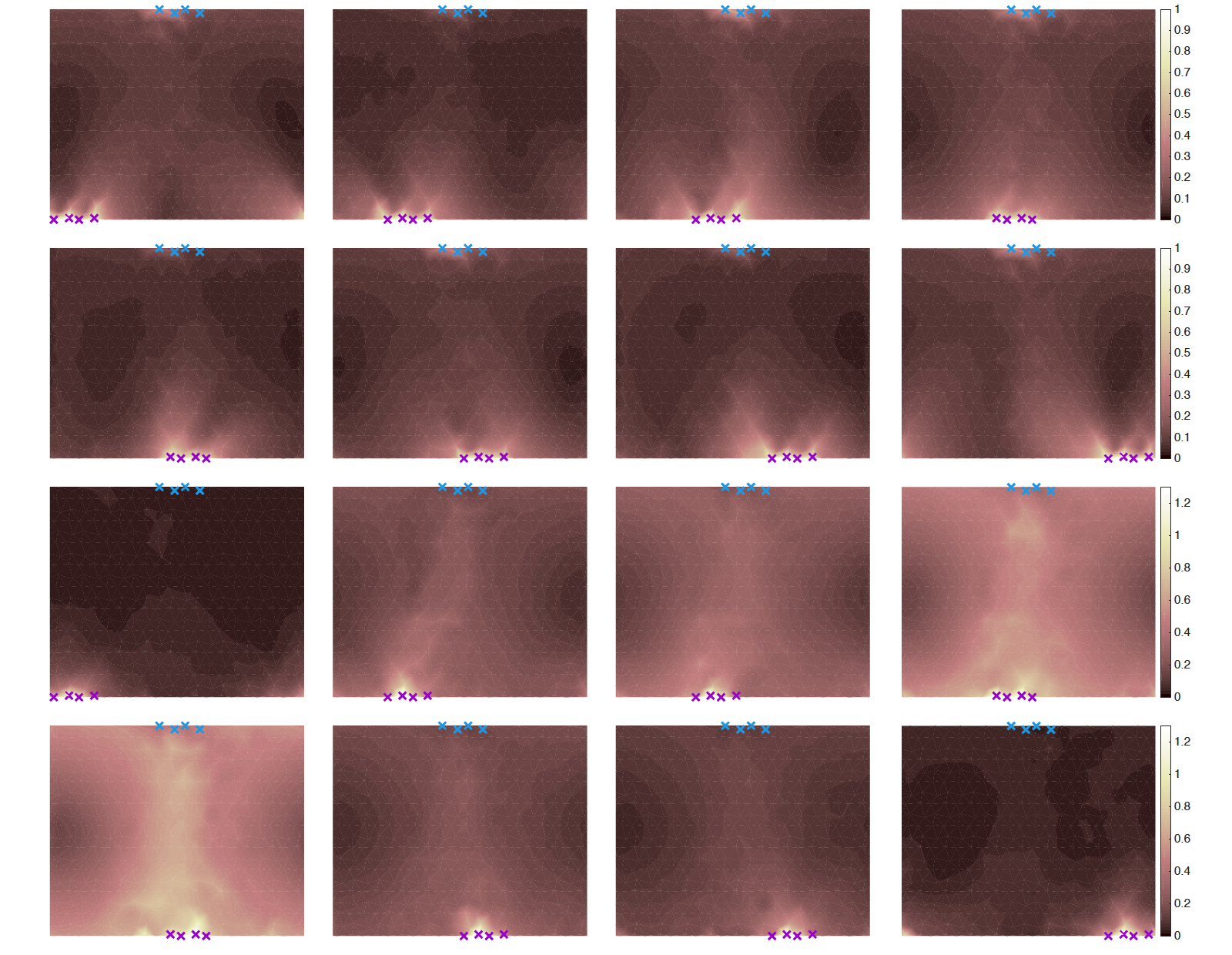}
\caption{\small{Specificity of the ({\it Top two}) geometric solutions and ({\it Bottom two}) cooperative solutions (with a periodic boundary) toward imposing a stimulus at the surface of the material, at a site shown by the purple crosses which differs from the allosteric site where these materials were trained to respond. Blue crosses indicate the active site. A large response at the active site is always found in the geometric design (due to the presence of a lever) but not for the cooperative design.}
}\label{sensitivity}
\end{figure*}

\begin{figure}[htbp]
\includegraphics[width=\linewidth]{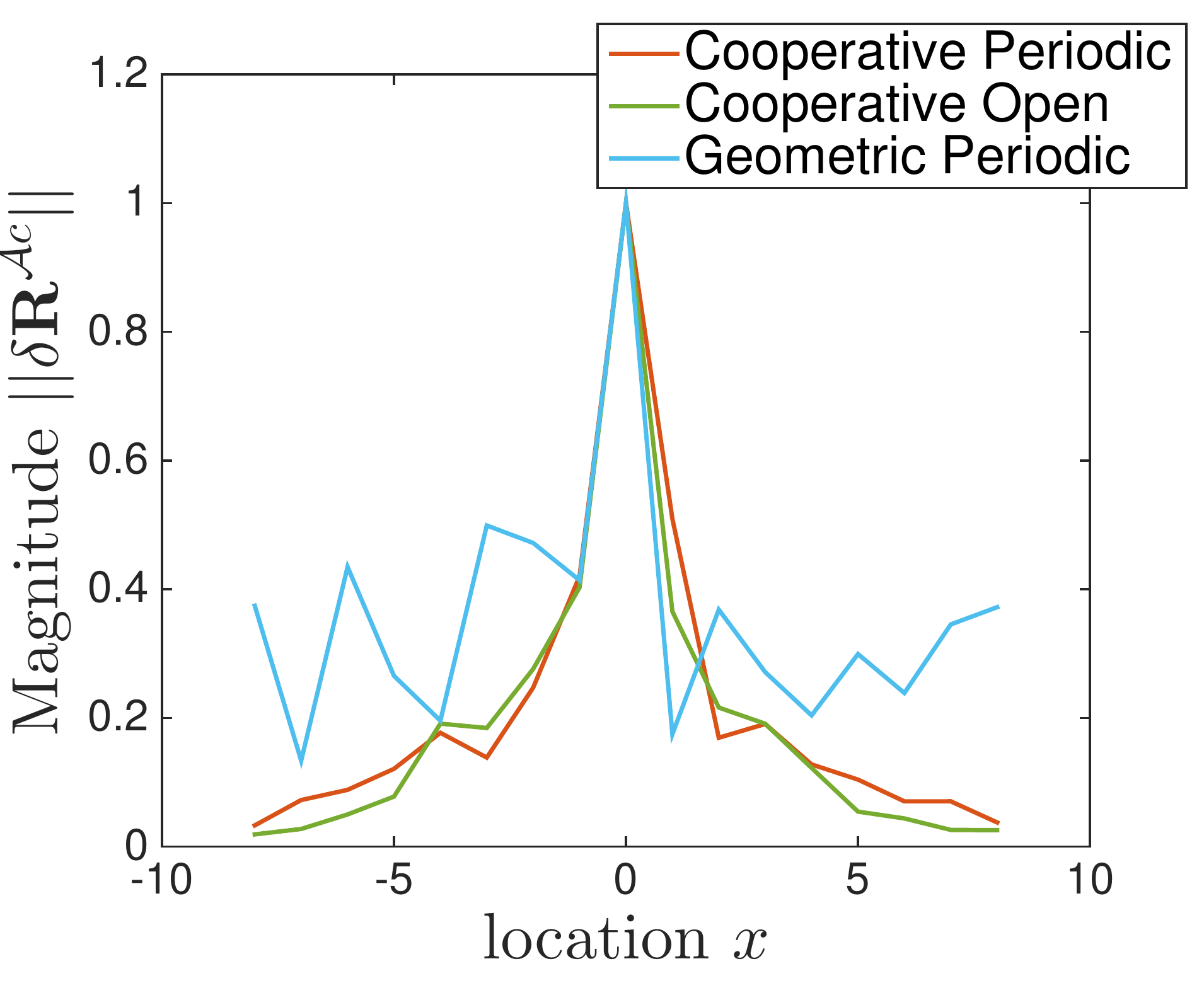}
\caption{\small{Average magnitude of response at the active site to a strain added on the other side of the system at location $x$ ($x=0$ corresponds to the position of the allosteric site) for both cooperative and geometric designs.}
}\label{sens_mag}
\end{figure}

{{\bf Specificity of the response:} Finally, in the geometric design a large response at the active site can be triggered by binding anywhere in the material, because the lever amplifies any elastic signal it finds, as shown  in the two upper panels of Fig.~\ref{sensitivity}. Thus geometric designs are not specific. By contrast, cooperative designs respond much more if the stimulus is triggered at the allosteric site (used to train the material), where the soft extended mode is designed to have a large shear, as shown  in the two lower panels of Fig.~\ref{sensitivity}. These results are quantified in Fig.\ref{sens_mag}.
}

\end{document}